\documentclass[conference]{IEEEtran}
\IEEEoverridecommandlockouts
\usepackage{mathptmx} % This is Times font

\usepackage{fancyhdr}
\usepackage[normalem]{ulem}
\usepackage[hyphens]{url}
\usepackage[sort]{cite}
\usepackage[final]{microtype}
\usepackage[bookmarks=true,breaklinks=true,letterpaper=true,colorlinks,citecolor=blue,linkcolor=blue,urlcolor=blue]{hyperref}

\usepackage{verbatim}
\usepackage{mathptmx} % This is Times font
\usepackage{amsmath,amssymb,amsfonts}
\usepackage{algorithm}
\usepackage[noend]{algpseudocode}
\usepackage{algorithmicx}
\usepackage{xcolor}
\usepackage{xspace}
\usepackage{xcolor, soul}
\sethlcolor{yellow}
\usepackage{booktabs}
\usepackage{siunitx}
\usepackage{multirow}
\usepackage{tabularx}
\usepackage{enumitem}
\usepackage{tikz}
\def\BibTeX{{\rm B\kern-.05em{\sc i\kern-.025em b}\kern-.08em
    T\kern-.1667em\lower.7ex\hbox{E}\kern-.125emX}}
\newcommand{\circlednumber}[1]{%
  \tikz[baseline=(char.base)]\node[shape=circle,draw,inner sep=0.7pt] (char) {#1};%
}

\newcommand{\paperN}{Strix\xspace}
\newcommand{\paperNcap}{STRIX\xspace}
\def\BibTeX{{\rm B\kern-.05em{\sc i\kern-.025em b}\kern-.08em
    T\kern-.1667em\lower.7ex\hbox{E}\kern-.125emX}}
\begin{document}

% Redefine the \title command to include a font size change
\let\oldtitle\title
\renewcommand{\title}[1]{\oldtitle{\huge #1}}

% Now use \title as usual
\title{\paperN: An End-to-End Streaming Architecture \\ with Two-Level Ciphertext Batching for Fully Homomorphic Encryption with Programmable Bootstrapping}

% \title{Conference Paper Title*\\

\author{
\IEEEauthorblockN{
Adiwena Putra,
Prasetiyo,
Yi Chen,
John Kim,
Joo-Young Kim}
\IEEEauthorblockA{\textit{School of Electrical Engineering} \\
\textit{KAIST}\\
Daejeon, South Korea \\
Email: \{adiwena.research, pras, chenyi, jjk12, jooyoung1203\}@kaist.ac.kr}
}

% \makeatletter
% \newcommand{\newlineauthors}{%
%   \end{@IEEEauthorhalign}\hfill\mbox{}\par
%   \mbox{}\hfill\begin{@IEEEauthorhalign}
% }
% \makeatother
% \author
% {
% \IEEEauthorblockN{\large Adiwena Putra}
% \IEEEauthorblockA{\textnormal{adiwena.research@kaist.ac.kr}\\
% \textnormal{KAIST}\\\textnormal{Daejeon, South Korea}}
% \and
% \IEEEauthorblockN{\large Prasetiyo}
% \IEEEauthorblockA{\textnormal{pras@kaist.ac.kr}\\
% \textnormal{KAIST}\\\textnormal{Daejeon, South Korea}}
% \and
% \IEEEauthorblockN{\large Yi Chen}
% \IEEEauthorblockA{\textnormal{chenyi@kaist.ac.kr}\\
% \textnormal{KAIST}\\\textnormal{Daejeon, South Korea}}
% \newlineauthors
% \IEEEauthorblockN{\large John Kim}
% \IEEEauthorblockA{\textnormal{jjk12@kaist.ac.kr}\\
% \textnormal{KAIST}\\\textnormal{Daejeon, South Korea}}
% \and
% \IEEEauthorblockN{\large Joo-Young Kim}
% \IEEEauthorblockA{\textnormal{jooyoung1203@kaist.ac.kr}\\
% \textnormal{KAIST}\\\textnormal{Daejeon, South Korea}}
% }

\maketitle

\begin{abstract}
Homomorphic encryption (HE) is a type of cryptography that allows computations to be performed on encrypted data. The technique relies on learning with errors problem, where data is hidden under noise for security. To avoid accumulating too much noise, the process of bootstrapping is needed to reset the noise level in the ciphertext, but it requires a large bootstrapping key and is computationally expensive. The fully homomorphic encryption over the torus (TFHE) scheme offers a faster and programmable bootstrapping (PBS) algorithm, which is crucial for security-focused applications like machine learning. Nonetheless, the current TFHE scheme does not support ciphertext packing, resulting in low-throughput performance. To the best of our knowledge, this is the first work that thoroughly analyzes TFHE bootstrapping, identifies the TFHE acceleration bottleneck in GPUs, and proposes a hardware TFHE accelerator to solve the bottleneck.

We begin by identifying the TFHE acceleration bottleneck in GPUs, which is caused by the blind rotation fragmentation problem. This issue can be significantly improved by increasing the batch size in PBS. We propose a two-level batching approach to substantially enhance the batch size in PBS. In order to efficiently implement this solution, we propose \paperN, which utilizes streaming and fully pipelined architecture with specialized functional units to accelerate the sequential ciphertext processing in TFHE. In particular, we propose a novel microarchitecture for decomposition in TFHE, suitable for processing streaming data at high throughput. We also utilize a fully-pipelined FFT microarchitecture to eliminate the complex memory access bottleneck and enhance its performance through a folding scheme, resulting in $2\times$ throughput improvement and $1.7\times$ area reduction. \paperN achieves over $1,067\times$ and $37\times$ higher throughput in running TFHE with PBS than the state-of-the-art implementation on CPU and GPU, respectively, outperforming the state of the art TFHE accelerator by $7.4\times$. 
\end{abstract}


\section{Introduction}
\label{sec:intro}

Homomorphic encryption (HE) allows data to be processed while data remains encrypted and provides significant improvement in privacy and security~\cite{rivest1978data}. One fundamental component of most HE schemes is that noise is introduced to maintain data security and limit the amount of computation that is possible. However, the introduction of bootstrapping enabled fully homomorphic encryption (FHE) where noise in the ciphertext is refreshed and enables an unlimited amount of computation on the data~\cite{gentry2009fully}. The benefits of FHE are significant, not only from a security and privacy perspective but also in enabling the secure offloading of computation to the cloud without revealing data. Unfortunately, one of the most significant challenges of FHE is the high computation cost. For instance, a baseline inference can take 100 ms on CPU, but an FHE implementation can take up to 3 hours, resulting in more than 100,000$\times$ slowdown~\cite{lee2022privacy}.

Recently, there has been significant effort in accelerating FHE, utilizing not only GPUs~\cite{jung2021over, shivdikar2022accelerating, dai2015cuhe, nufhe} and FPGAs~\cite{riazi2020heax, agrawal2022fab, yang2023poseidon, gener2021fpga}, but also dedicated hardware accelerators~\cite{samardzic2021f1, samardzic2022craterlake, kim2022bts, kim2022ark}. Most prior work on acceleration has focused on the popular FHE scheme, CKKS~\cite{cheon2017homomorphic}, which supports floating-point representation and is suitable for machine learning workloads. However, while CKKS is powerful, it is still costly both in terms of computation and cost. For example, a recent accelerator for CKKS delivers significant performance improvement, but its estimated cost is approximately 418 mm$^2$ (nearly equivalent to a high-end GPU~\cite{kim2022ark}) with 512 MB of on-chip memory. 

In this work, we investigate a different type of FHE based on Fully Homomorphic Encryption over the Torus (TFHE) \cite{chillotti2020tfhe}, which offers lower computational and memory requirements compared to other FHE schemes. Similar to other FHE schemes, the most time-consuming operation in TFHE is bootstrapping, which is necessary to refresh noise during homomorphic computations. However, a crucial advantage of TFHE bootstrapping is its programmability, allowing the evaluation of any univariate function while performing bootstrapping simultaneously through programmable bootstrapping (PBS). The computational requirements, as well as the input data structure, 
% and \hl{memory access pattern} 
fundamentally differ from other FHE schemes like CKKS. Consequently, prior work on FHE accelerators~\cite{samardzic2021f1, samardzic2022craterlake, kim2022bts, kim2022ark, agrawal2022fab, yang2023poseidon} are not directly applicable to TFHE.

The primary challenge in TFHE is achieving high throughput programmable bootstrapping (PBS), as it can only be executed for one message at a time. Our analysis reveals that the main bottleneck in PBS is the blind rotation, which involves multiple iterations of vector-matrix multiplication of polynomial that must be performed sequentially. To improve PBS performance, either the latency or throughput of PBS can be addressed. To improve PBS latency, a single blind rotation can be accelerated when processing one ciphertext (i.e., accelerate individual iteration of blind rotation). When there are multiple ciphertexts, blind rotation needs to be performed multiple times for each ciphertext~\cite{jiang2022matcha, ye2022HPEC}. In comparison, PBS throughput can be improved by enabling a single blind rotation to process multiple ciphertexts simultaneously~\cite{nam22nbit, nufhe}.
While reducing the latency of PBS improves performance, the workload becomes memory-bound because of a shorter time gap to fetch the next bootstrapping key for the next blind rotates iteration. In comparison, addressing the throughput of PBS becomes compute bound as more temporal reuse of bootstrapping key, resulting in a larger time gap to fetch the next bootstrapping key. However, in this work, we analyze PBS on GPUs and identify how attempting to improve throughput can result in \emph{blind rotation fragmentation} where multiple blind rotations need to be performed sequentially and significantly deteriorate performance.

To address blind rotation fragmentation, increasing batch size of a single blind rotation is essential, which is often constrained by the number of cores. We introduce the concepts of device-level and core-level batching to significantly expand the batch size for a single blind rotation. Device-level ciphertext batching substantially enhances spatial parallelism by enabling each core to work on different ciphertexts while utilizing the same set of bootstrapping keys, thereby increasing spatial reuse. Conversely, core-level batching improves temporal parallelism by allowing each core to process a continuous stream of ciphertexts using the same set of bootstrapping keys. However, efficiently implementing core-level batching necessitates specialized hardware units designed to exploit available parallelism and amortize the cost of processing ciphertexts sequentially in each blind rotation iteration.

We introduce \paperN, a hardware accelerator designed to achieve high throughput TFHE by leveraging two-level batching to significantly increase the batch size of a single blind rotation, ultimately mitigating blind-rotate fragmentation problem. \paperN is equipped with five functional units to efficiently handle the compute workload in one blind rotation iteration. To determine the optimal balance among our functional units, we performed a comprehensive analysis of the PBS algorithm and discovered four distinct levels of parallelism in PBS. This analysis laid the foundation for designing the \paperN accelerator architecture, allowing for effective pipelining with nearly 100\% utilization and maximum throughput.

Among these functional units, the decomposer unit and (I)FFT unit play critical roles and present significant design challenges. The decomposer unit's task is to decompose one polynomial into smaller polynomials. We present a novel microarchitecture for the decomposer unit that effectively consumes a stream of polynomials and produces a decomposed stream of polynomials with constant throughput, using efficient masking and buffering techniques. Another important functional unit is the I/FFT unit. Unlike the I/FFT units implemented in CKKS accelerators (I/NTT in their case), which employ 2D NTT requiring matrix transposition between different steps, we adopt a parallel pipelined FFT unit that processes a constant stream of polynomials without the need for matrix transposition. This design aligns well with our streaming architecture. We further optimize the FFT unit with a folding scheme technique, allowing the transformation of N-point polynomials using an N/2-point FFT unit, effectively reducing the latency and hardware cost by 50\%. 

We implement all functional units in a single Homomorphic Streaming Core (HSC). By integrating up to 8 HSCs on a single silicon chip and carefully orchestrating the data flow, we achieve an impressive 1,067$\times$ and 37$\times$ increase in throughput, and a speedup of $38\times$ and $17\times$ for machine learning workloads, respectively, when compared to state-of-the-art CPU and GPU implementations.

In this paper, we make the following key contributions:
\begin{itemize}
    \item We perform an in-depth analysis of TFHE programmable bootstrapping (PBS) on a GPU and identify a critical performance issue stemming from the blind rotation fragmentation problem. This issue arises because the blind rotation process must be executed multiple times, leading to significant performance degradation.
    \item We propose \paperN, a hardware accelerator designed to mitigate the blind rotation fragmentation problem through the concepts of device-level and core-level batching to substantially enhance the batch size for a single-blind rotation.
    \item To efficiently implement core-level batching, specialized hardware units are required to exploit any available parallelism and amortize the cost of sequentially processing ciphertexts in each blind rotation iteration. We perform a comprehensive analysis of PBS parallelism, identifying four distinct levels of parallelism that serve as the foundation for designing the \paperN accelerator architecture.
    \item We introduce a novel microarchitecture for \paperN, featuring five interconnected functional units that achieve nearly 100\% utilization. We present an innovative decomposer unit that efficiently processes polynomial streams and a fully pipelined FFT unit that eliminates matrix transposition and incorporates a folding scheme for enhanced blind rotation computation.
\end{itemize}

\section{Background}
\label{background}
\label{sec:background}
\subsection{Homomorphic Encryption}

Homomorphic Encryption (HE) is a cryptographic technique that enables computations to be performed on encrypted data without decryption~\cite{rivest1978data}. Modern HE schemes rely on lattice problems, which are known to be secure against quantum computers \cite{regev2009lattices,bernstein2009introduction,khot2005hardness}. Learning With Errors (LWE), a lattice-based cryptography, is commonly used to implement HE systems \cite{gentry2009fully} as LWE  ensures data security by concealing it under noise. However, as computations are performed on encrypted data, the noise accumulates, and if the noise exceeds a certain threshold, the original data cannot be decrypted. Fully Homomorphic Encryption (FHE)  overcomes this limitation with bootstrapping mechanism that can refresh the noise in the ciphertext back to its initial level \cite{gentry2009fully}. Unfortunately, bootstrapping is computationally expensive and requires minutes to hours of execution for a single ciphertext \cite{gentry2009fullybook,gentry2011implementing}. As a result, different HE schemes have been proposed to improve the efficiency of bootstrapping~\cite{brakerski2014leveled,fan2012somewhat,cheon2017homomorphic,gentry2013homomorphic,ducas2015fhew,chillotti2016faster}.

HE schemes are typically classified into two categories: BGV-like and GSW-like schemes \cite{falcetta2022privacy, podschwadt2022survey, marcolla2022survey}. BGV-like schemes (e.g., BGV \cite{fan2012somewhat}, BFV \cite{brakerski2014leveled}, and CKKS \cite{cheon2017homomorphic}) encrypt messages at the vector level, with every homomorphic operation being a vector operation. Although bootstrapping in BGV-like schemes is relatively slow, they offer high throughput bootstrapping. In contrast, GSW-like schemes (e.g., GSW\cite{gentry2013homomorphic}, FHEW \cite{ducas2015fhew}, and TFHE \cite{chillotti2016faster}) encrypt messages at the word or bit level, offering greater flexibility to support a wide range of operations, including arithmetic, logical, and relational operations. This flexibility is achieved through functional or programmable bootstrapping (PBS), which can perform homomorphic look-up tables. The higher expressive power of GSW-like schemes is critical for HE to handle diverse applications. 

\subsection{TFHE Overview}

Fast fully homomorphic encryption scheme over the torus (TFHE) was proposed to enable bootstrapping to be performed in under 0.1 seconds using a single-core CPU \cite{chillotti2016faster}. Unlike prior encryption schemes~\cite{fan2012somewhat, brakerski2014leveled, cheon2017homomorphic} that require separate (dedicated) bootstrapping steps to reset noise, TFHE supports programmable bootstrapping which allows bootstrapping to be performed during the evaluation of arbitrary univariate functions \cite{chillotti2021programmable}. TFHE was initially limited to performing boolean algebra operations but has been extended to include operations for integer and fixed-point numbers \cite{chillotti2016homomorphic,chillotti2017faster,chillotti2020tfhe,chillotti2021improved,chillotti2021programmable,chillotti2022scooby}. Homomorphic addition and programmable bootstrapping are the two fundamental homomorphic operations supported by TFHE, which can be used for constructing any function, including logical and relational operations. In general, PBS enables the construction of any function of one variable, making TFHE highly versatile for a wide range of homomorphic computation applications.

\subsection{Alternative FHE Schemes}
\begin{table}[t]
\centering
\caption{Comparison between CKKS and TFHE}
\label{table:ckks_vs_tfhe}
\footnotesize
\begin{tabular}{@{}lll@{}}
\toprule
\textbf{} & \multicolumn{1}{c}{\textbf{CKKS}} & \multicolumn{1}{c}{\textbf{TFHE}} \\
\midrule
\textbf{Ciphertext size} & MB level & KB level \\
\textbf{Bootstrapping key size} & GB level & 10s-100s MB level \\
\textbf{Computational bottleneck} & NTT & FFT/NTT \\
\textbf{Ciphertext packing} & supported & not supported \\
\textbf{Bootstrapping latency} & high latency & low latency \\
\textbf{Bootstrapping throughput} & high throughput & low throughput \\
\textbf{Supported operations} & Add, multiply & Add, look-up table \\
\textbf{Message type} & real, complex & real, integer, boolean \\
\bottomrule
\end{tabular}
\end{table}

One of the popular FHE schemes is the CKKS algorithm~\cite{cheon2017homomorphic}, which is known for its ability to support floating-point numbers. Many recent works have addressed the computation challenges in CKKS, including various microarchitectural explorations and accelerators~\cite{samardzic2022craterlake, kim2022bts, kim2022ark, yang2023poseidon, agrawal2022fab}. While these works provide significant performance improvements compared to naive implementations on CPUs or GPUs, the performance gap between FHE and native workloads remains substantial, particularly due to significant bootstrapping overhead. In comparison, this work targets TFHE, which has fundamentally different requirements and trade-offs from CKKS. A high-level qualitative comparison between TFHE and CKKS is summarized in Table~\ref{table:ckks_vs_tfhe}.
From a computational standpoint, TFHE is more lightweight than CKKS, with considerably smaller ciphertext and bootstrapping key sizes, making it more practical for low to mid-range computing systems. Both CKKS and TFHE have computational bottlenecks in polynomial transformations for polynomial multiplication. However, TFHE offers more flexibility as it allows the use of FFT or NTT for polynomial transformation, which opens up more algorithm optimization possibilities.

Another significant difference is in bootstrapping. Bootstrapping represents a significant fraction of the overall execution time for both CKKS and TFHE. Although bootstrapping in TFHE is considerably faster than CKKS, TFHE only allows for bootstrapping one message at a time. On the other hand, CKKS allows for bootstrapping a vector of messages at a time due to ciphertext packing capability, resulting in higher throughput performance. However, TFHE offers more diverse operations due to PBS, which can evaluate any function of single variable. This unique feature of TFHE schemes allows them to evaluate logical and relational operations, which are difficult to perform in the CKKS scheme.

From application perspective, TFHE is particularly useful for evaluating the activation function in neural networks \cite{lou2019she, chillotti2021programmable, stoian23dnntfhe}, evaluating tree-based machine learning models \cite{frery23pptree}, or emulating the CPU, which can run encrypted programs \cite{matsuoka2021virtual}. Overall, while CKKS is well-suited for homomorphic encryption where large numbers of the real/complex number are involved, TFHE's versatility in its message type and support for PBS makes it a practical choice for general-purpose homomorphic encryption applications.

\subsection{Data Structure in TFHE}

\begin{table}[t]
\normalsize
\centering
\caption{Typical parameter value for TFHE scheme}
\label{table:tfhe_param}
\begin{tabular}{@{}lll@{}}
\toprule
\textbf{Parameter} & \textbf{Description} & \textbf{Typical value} \\
\midrule
$n$ & LWE mask & 256--4096 \\
$N$ & Polynomial degree & $2^{10}$--$2^{14}$ \\
$k$ & GLWE mask & 1--4 \\
$l_b$ & decomposition level & 2--4 \\
\bottomrule
\end{tabular}
\end{table}

Entities involved in PBS and keyswitching operations in TFHE can be categorized into four groups: LWE ciphertext, GLWE test-vector, bootstrapping key (bsk), and keyswitching key (ksk). LWE ciphertext is represented as a vector containing $(n+1)$ scalar elements $[a_1,...,a_{n},b]$, where each scalar element is a 32-bit or 64-bit integer. This ciphertext type is primarily used to encrypt messages in TFHE. On the other hand, GLWE is a ciphertext type that stores univariate functions for PBS. It is represented as a vector of $(k+1)$ $N-1$ degree polynomials $[A_1(X),...,A_k(X),B(x)]$, where each polynomial is represented as a vector of $N$ integers, either 32-bit or 64-bit.

The bootstrapping key and keyswitching key serve as "parameters" that operate with ciphertext and test-vector during PBS or keyswitching operations. The bootstrapping key is a vector of $n$ General-GSW (GGSW) ciphertexts, where each GGSW ciphertext is a $(k+1) \cdot l_b \times (k+1)$ matrix of $N-1$ degree polynomials. On the other hand, the keyswitching key is a vector of $k\cdot N\cdot l_k$ LWE ciphertexts. Table~\ref{table:tfhe_param} shows the typical parameter values for the TFHE scheme.

In summary, LWE and GLWE represent the encrypted message and the univariate function, respectively, while the bootstrapping key and keyswitching key serve as the parameters to operate with the ciphertext and test-vector during PBS or keyswitching operations. These entities play a vital role in the homomorphic encryption operations in the TFHE scheme and their efficient implementation is crucial for the performance of the scheme. For more details of the above entities, refer to the original cryptographic algorithm paper \cite{chillotti2017faster}.

\begin{algorithm}[t]
\small
\caption{\small{Programmable Bootstrapping}}
\label{algorithm_pbs}
\begin{algorithmic}[1]
\Require{$n,k$: mask length in LWE/GLWE ciphertext}
\Require{$l_b$: decomposition level of bootstrapping}
\Require{$N$: number of coefficients in polynomial}
\Require{$c[n+1]$: LWE ciphertext, where each element is scalar}
\Require{$tv[k+1]$: test vector, each element is N degree polynomial}
\Require{$bsk[n][(k+1)l_b][k+1]$: bootstrapping key, each element is N degree polynomial}
\Ensure{$o[kN+1]$: LWE ciphertext, where each element is scalar}
\State{\textbf{Local variables:} $etv[(k+1)l_b]$, $otv[k+1]$ (initialized to zero)}
\For{$i = 0$ \textbf{to} $(n+1)$}
\State{$c[i]$ = ModSwitch$(c[i])$} \Comment{\textcolor{olive}{// Modulus Switching}}
\EndFor
\State{$tv$ = Rotate('left', $tv$, $c[n]$)} 
\For{$i = 0$ \textbf{to} $n$ } \Comment{\textcolor{olive}{// Blind Rotation}}
\State{$tv$ = $tv$ - Rotate('Right', $tv$, $c[i]$)} \Comment{\textcolor{olive}{// Rotate and subtract}}
\State{$etv$ = Decompose($tv$, $l_b$)} \Comment{\textcolor{olive}{// Decomposition}}
\For{$j = 0$ \textbf{to} $(k+1)$ }
\For{$k = 0$ \textbf{to} $(k+1)l_b$ }
\State{$otv[j]$=$otv[j]$+IFFT(FFT($etv[k]$) * FFT($bsk[i][k][j]$))} \Comment{\textcolor{olive}{// External product}}
\EndFor
\EndFor
\State{$tv = otv$}
\State{initZero($otv$)}
\EndFor
\State{$o$ = SampleExtract($tv$)} \Comment{\textcolor{olive}{// External Sample Extract}}
\end{algorithmic}
\end{algorithm}

\begin{algorithm}[t]
\small
\caption{\small{Keyswitching}}
\label{algorithm_keyswitch}
\begin{algorithmic}[1]
\Require{$n,k$: input mask length of LWE/GLWE ciphertext}
\Require{$l_k$: decomposition level of keyswitching}
\Require{$c[n+1]$: LWE ciphertext, where each element is scalar}
\Require{$ksk[nl_k][m+1]$: keyswitching key, each element is scalar}
\Ensure{$o[m+1]$: LWE ciphertext, where each element is scalar (initialized to zero)}
\State{\textbf{Local variable:} $eic[nl_k]$ (initialized to zero)}
\State{$o[m] = c[n]$} \Comment{\textcolor{olive}{// Decomposition }}
\State{$eic$ = Decompose($c[0..n]$, $l_k$)}
\For{$i = 0$ \textbf{to} $m+1$}
\For{$j = 0$ \textbf{to} $n*l_k$}
\State{$o[i] = o[i] - eic[j]*ksk[j][i]$} \Comment{\textcolor{olive}{// Vector matrix multiplication }}
\EndFor
\EndFor
\end{algorithmic}
\end{algorithm}

\subsection{PBS and Keyswitching}

\textbf{Programmable bootstrapping (PBS).}
Algorithm \ref{algorithm_pbs} outlines the PBS procedure in TFHE. To enhance its readability, we have presented the original algorithm~\cite{chillotti2020tfhe, joye2022sok} in the form of matrix and vector operations. PBS is composed of three building blocks: modulus switching, blind rotation, and sample extract. During modulus switching, each element in LWE is switched from the original modulus $q$ to the new modulus $2N$. This operation is inexpensive and straightforward since $N$ is always power of two integers. Blind rotation, which constitutes 96\% of the PBS execution time, is the core operation of PBS.  Blind rotation, the core PBS operation, constitutes 96\% of its execution time and involves non-parallelizable iterations, each with three steps: polynomial rotation and subtraction, decomposition, and external product. Lastly, sample extract forms the new LWE ciphertext by extracting a specific coefficient from the GLWE.

\textbf{Keyswitching.} 
After performing PBS, the resulting LWE ciphertext is encrypted under a different key. It becomes a vector of $k\cdot N + 1$ scalar elements. To decrypt it using the original key, keyswitching algorithm is needed to convert the ciphertext back to its original size (a vector of $n+1$ scalar elements). Algorithm~\ref{algorithm_keyswitch} summarizes the keyswitching algorithm~\cite{chillotti2020tfhe, joye2022sok}.
Keyswitching begins by decomposing the input LWE ciphertext, taking the first $k\cdot N$ elements in LWE and extending it into a $k \cdot N \cdot l_k$-element vector, where $l_k$ is the decomposition level for keyswitching. This vector is then multiplied by a $k \cdot N \cdot l_k \times n + 1$ matrix, resulting in a vector of $n+1$ elements that represent the new LWE ciphertext encrypted under the original key.
\section{Motivation: Identifying Bottleneck in TFHE}
\label{sec:motivation}

%%%%%%%%%%%%%%%%%%%%%%%%%%%%%%%%%%% FIGURE %%%%%%%%%%%%%%%%%%%%%%%%%%%%%%%%%%% 
\begin{figure}[t]
    \centering
    \includegraphics[width=\linewidth]{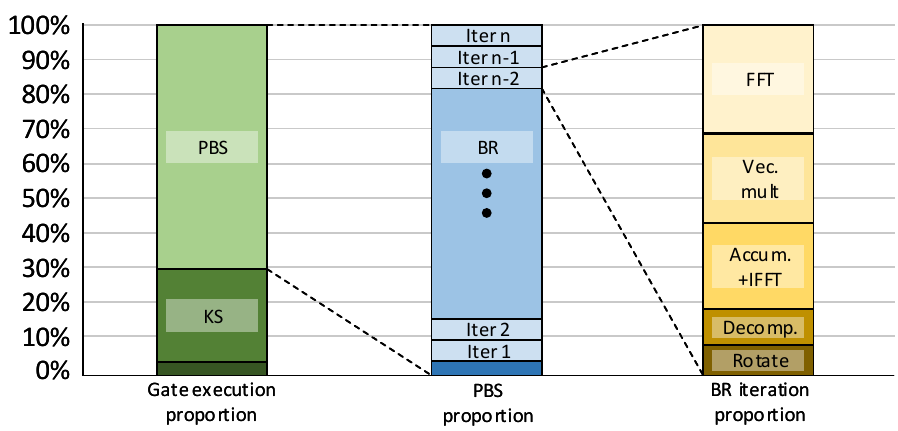}
    \caption{Workload breakdown for TFHE gate operation on CPU.}
    \label{fig:proportion}
\end{figure}

The main bottleneck in TFHE is caused by PBS operations which do not support packing, so each message needs to be bootstrapped one by one at a time, eventually resulting in low throughput PBS. Thus, it is clear that the primary goal of TFHE accelerator is to increase the PBS throughput. To accomplish this, we first conduct a workload breakdown analysis of TFHE performance on a CPU to determine the overall workload distribution. Next, we perform an initial study of TFHE performance on a GPU to identify the main bottlenecks when accelerating TFHE on commercial hardware accelerators. We use the Concrete library \cite{chillotti2020concrete} for TFHE workload breakdown. For the GPU implementation of TFHE, we use the NuFHE library \cite{nufhe}, and utilize Nvidia's Nsight profiling tools to obtain kernel execution performance metrics on an Nvidia Titan RTX GPU.

Figure~\ref{fig:proportion} displays the breakdown of TFHE gate operation and execution time distribution on a CPU. At the highest level of the call stack, approximately 65\% of the computation time is spent on PBS, 30\% on keyswitching (KS), and the remaining 5\% on other operations such as homomorphic addition/subtraction. The execution time proportion may vary slightly depending on the chosen security parameter. In our experiment, we utilize the default 110-bit security parameter from the NuFHE library in the concrete library. Within the PBS function, approximately 98\% of the execution time is dedicated to blind rotation (BR), while the remaining time is allocated to modulus switching and sample extraction. The blind rotation consists of sequential iterations, where the majority of each iteration's execution time is consumed by the external product. This external product can be broken down into three parts: FFT, vector multiplier, and IFFT. IFFT processes less number of polynomials than FFT due to the decomposition operation, resulting in a shorter execution time.
The ratio of the number of polynomials processed by FFT and IFFT is $l_b : 1$, making the workload imbalance more severe as $l_b$ increases.

The most time-consuming operation of blind rotation is the external product which performs matrix-vector multiplication with polynomials as elements. This process is executed for every ciphertext. When multiple ciphertexts are considered, one can envision multiple nodes, each performing blind rotation for its respective ciphertext. Since the bootstrapping key (a matrix of polynomials) can be shared among nodes (provided they all work on the same blind rotation iteration), it is possible to batch multiple ciphertexts and perform matrix-matrix multiplication with polynomials in each iteration. This is precisely what we observe in NuFHE implementations when the optimal security parameter was chosen. All GPU cores (SM) work on the same blind rotation iteration, allowing them to share a portion of the bootstrapping key. We refer to this strategy as \textbf{\textit{device-level batching}}: each core works on a different ciphertext's PBS but all cores share the same bootstrapping key in the same iteration of blind rotation.

%%%%%%%%%%%%%%%%%%%%%%%%%%%%%%%%%%% FIGURE %%%%%%%%%%%%%%%%%%%%%%%%%%%%%%%%%%% 
\begin{figure}[t]
    \centering
    \includegraphics[width=\linewidth]{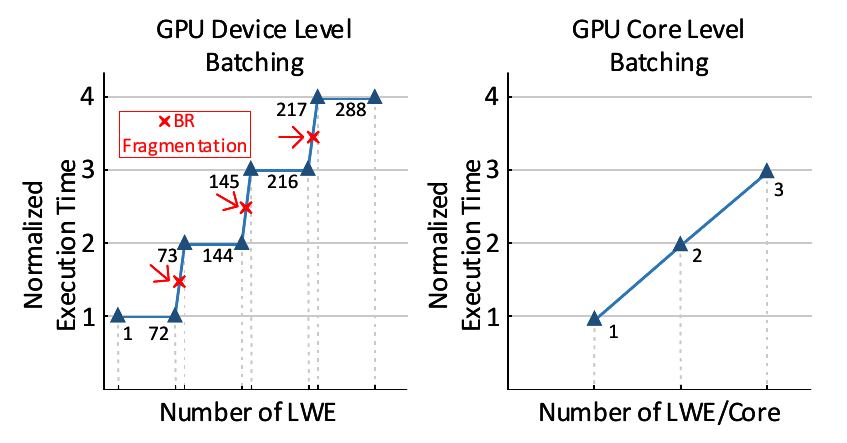}
    \caption{Execution time of the blind rotation kernel on a GPU, illustrating the fragmentation caused by device-level batching and the corresponding increase in execution time when core-level batching is implemented on the GPU.}
    % \vspace{-0.1in}
    \label{fig:nufhe}
\end{figure}

Figure~\ref{fig:nufhe} displays the profiling results for the blind rotation kernel in the NuFHE library. Due to the device-level batching, the execution time is kept constant for the first 72 ciphertexts, as our GPU has 72 cores. Nevertheless, when the number of ciphertexts surpasses the available cores (i.e., all cores are fully utilized), blind rotation (BR) fragmentation occurs, resulting in a twofold increase in the overall execution time for the blind-rotation kernel. The TFHE execution time on a GPU can be expressed as:
\begin{equation}
    \text{Total time} = \text{(\# BR fragmentations + 1)} \times \text{BR time per core}
\end{equation}
where the number of BR fragmentations is determined by:
\begin{equation}
    \text{\# BR fragmentations} = \left\lceil\frac{\text{\# ciphertexts}}{\text{batch size}}\right\rceil - 1
\end{equation}

To boost overall throughput, we can either reduce the number of BR fragmentations or decrease the BR execution time per core. Reducing the BR execution time can be achieved by exploiting parallelization within each blind rotation iteration. On the other hand, reducing the fragmentation number can be accomplished by batching more ciphertexts in a single BR fragment. In the case of NuFHE, the number of ciphertexts that can be batched is always fixed to the number of cores. Increasing the number of cores is effective, but it is costly and not scalable. This consideration leads us to the concept of \textbf{\textit{core-level batching}}, wherein multiple ciphertexts can be batched into a single core, effectively increasing the batch size of a single BR execution to the product of core-level batch size and device-level batch size.

Utilizing core-level batching means allocating multiple ciphertexts to each core, requiring the sequential processing of multiple ciphertexts in a single-blind rotation iteration.
In the case of GPUs, even if we reduce the number of BR fragmentation with core-level batching, the total execution time does not decrease at all as the execution time for each iteration in blind rotation increases accordingly, as illustrated in Figure~\ref{fig:nufhe}. 
This motivates us to design a specialized TFHE core that can efficiently process a series of ciphertexts. With customized and fully pipelined arithmetic datapath, we can accelerate the sequential ciphertext processing in a core and amortize the hardware cost. Overall, multiple cores process many ciphertexts in parallel, while each core accelerates sequential ciphertext processing with specialized logic.

Besides PBS, another major time-consuming operation in TFHE is the keyswitching operation. Similar to PBS, keyswitching in TFHE is performed individually for each ciphertext. Our measurements indicate that the execution time for keyswitching accounts for approximately 30\% of the total execution time. In the case of NuFHE, keyswitching is accelerated using different kernels from those for blind rotation, introducing additional latency. By designing a specialized hardware unit, we can hide the latency arising from the keyswitching operation, which is commonly executed after PBS. Therefore, developing a dedicated hardware accelerator tailored to address these challenges will significantly enhance the efficiency and performance of TFHE operations.
\section{\paperNcap Architecture}

\label{architecting}
%In this section, we propose the \paperN architecture for TFHE acceleration. We begin by outlining our parallelism strategy, which establishes the foundation for the overall architecture. Next, we describe the \paperN architecture, including its organization and functional units. Finally, we explain the workload scheduling and compute pipeline in \paperN.

%%%%%%%%%%%%%%%%%%%%%%%%%%%%%%%%%%% FIGURE %%%%%%%%%%%%%%%%%%%%%%%%%%%%%%%%%%% 
\begin{figure}[t]
    \centering
    \includegraphics[width=3.4in]{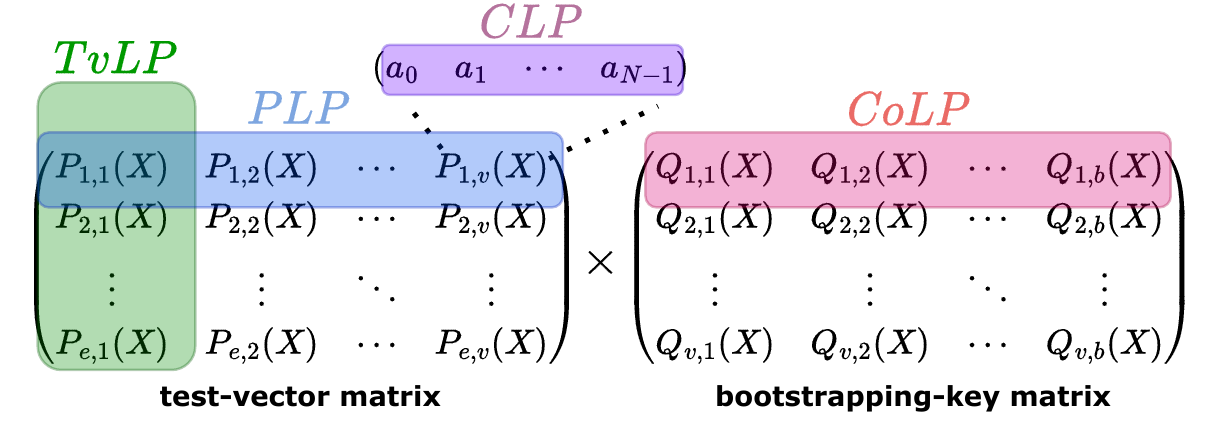}
    % \vspace{-0.3in}
    \caption{Illustration of four levels of parallelism when performing external product.}
    \label{fig:parallelism}
    %\vspace{-0.1in}
\end{figure}

\subsection{Parallelism Strategy}

As discussed in Section~\ref{sec:motivation}, the blind rotation process involves multiple iterations that cannot be parallelized. In each iteration, matrix-vector multiplications of polynomials must be performed for processing one LWE. However, if we consider processing multiple LWEs in a single iteration, i.e., input batching, the operation would become a polynomial matrix-matrix multiplication, which is the major workload for the accelerator. 
Within this notation, we discover four parallelism levels that can be leveraged in the \paperN architecture.

\textbf{Four parallelism levels in PBS.}
Figure~\ref{fig:parallelism} illustrates the four ways to exploit parallelism in matrix-matrix multiplication of polynomials. The most straightforward parallelism to leverage is test-vector level parallelism (\textit{TvLP}), which is the number of rows that can be processed concurrently in the test-vector matrix. For each row in the test-vector matrix, it is essential to consider the number of polynomials that can be processed in parallel, referred to as polynomial level parallelism (\textit{PLP}). Moreover, for each polynomial, the number of coefficients that can be processed in parallel, known as coefficient level parallelism (\textit{CLP}), must be considered. Finally, the number of columns in the output matrix that can be processed in parallel, called column level parallelism (\textit{CoLP}), should also be taken into account.

\textbf{Hardware implication on parallelism levels.}
Although all four levels of parallelism can be utilized simultaneously, each level has different implications for hardware implementation and cost. Exploiting excessive parallelism at a certain level can lead to expensive hardware cost. For example, \textit{TvLP} is same as the number of processing cores in the architecture, \textit{CLP} is associated with the number of lanes the computing unit, and \textit{PLP} and \textit{CoLP} are related to the replication the computing unit as discussed in Section~\ref{sec:micro}. 

\textbf{Parallelism availability.}
The degree of parallelism that can be achieved for $PLP$ and $CoLP$ is determined by the size of the vector or matrix utilized in the external product calculation. The maximum number of polynomials that can be processed in parallel for $PLP$ is $(k+1) \times l_b$, while the maximum number for $CoLP$ is $(k+1)$. Based on the latest TFHE library \cite{chillotti2020concrete}, the typical range of $k$ is between $1$ and $4$, while the typical range of $l_b$ is between $2$ and $4$. Therefore, we can expect a minimum parallelism of $PLP = 4$ and $CoLP=2$. The availability of parallelism for the $CLP$ can be achieved when performing FFT on an $N-1$ degree polynomial. The practical range for $N$ lies between $1,024$ and $16,384$. In the case of $TvLP$, parallel processing can only be utilized when executing the PBS operation on multiple ciphertexts. The degree of parallelism for $TvLP$ is dependent on the number of inputs, which varies by application. For instance, in a machine learning context, the image size of the MNIST dataset is $28 \times 28$, allowing for parallel execution of $784$ PBS, resulting in a maximum $TvLP$ value of $784$.

\textbf{Prioritization of the parallelisms.}
In order to create an efficient hardware accelerator, it is important to set the design goal and focus on it (e.g., whether it targets reducing latency or increasing throughput). 
We believe that prioritizing throughput is more important in the TFHE accelerator, as many applications require multiple ciphertexts PBS. To this end, we prioritize parallelism in the order of \textit{TvLP-CLP-PLP-CoLP}, taking into account parallelism availability, hardware cost, and throughput goals. \textit{TvLP} and \textit{CLP} offer more significant parallelism opportunities than \textit{PLP} and \textit{CoLP}, based on common security parameters and applications. When choosing between \textit{TvLP} and \textit{CLP}, we consider bandwidth and implications for the hardware. Increasing \textit{TvLP} involves adding more cores to process more LWEs in one epoch and increase throughput. Increasing \textit{CLP} also improves throughput performance, however it also results in a need for more external bandwidth due to the shorter duration between blind rotation iterations, which requires the bootstrapping key to be retrieved more frequently. Finally, considering the typical value of TFHE parameters and implementation feasibility, such as area, bandwidth, and interconnect cost, we chose $TvLP=8$, $CLP=4$, $PLP=2$, and $CoLP=2$.

\textbf{Parallelism in Keyswitching.}
When dealing with multiple ciphertexts, keyswitching becomes a matrix-matrix multiplication. Our strategy for parallelism in keyswitching is similar to that of external products, except without polynomial level parallelism ($PLP=1$) as we are no longer working with polynomials. For keyswitching processing, we use $TvLP=8$, $CLP=8$, and $CoLP=8$ in our \paperN architecture. The $TvLP$ value is determined by the number of cores, with the goal of having one keyswitch's functional unit cluster on each core. Meanwhile, $CLP$ is set to match the throughput of functional units from PBS computation. Finally, $CoLP$ is set as large as possible to utilize all channels from HBM fully. 

%%%%%%%%%%%%%%%%%%%%%%%%%%%%%%%%%%% FIGURE %%%%%%%%%%%%%%%%%%%%%%%%%%%%%%%%%%% 
\begin{figure*}[t] 
\centering
\footnotesize
\includegraphics[width=\linewidth]{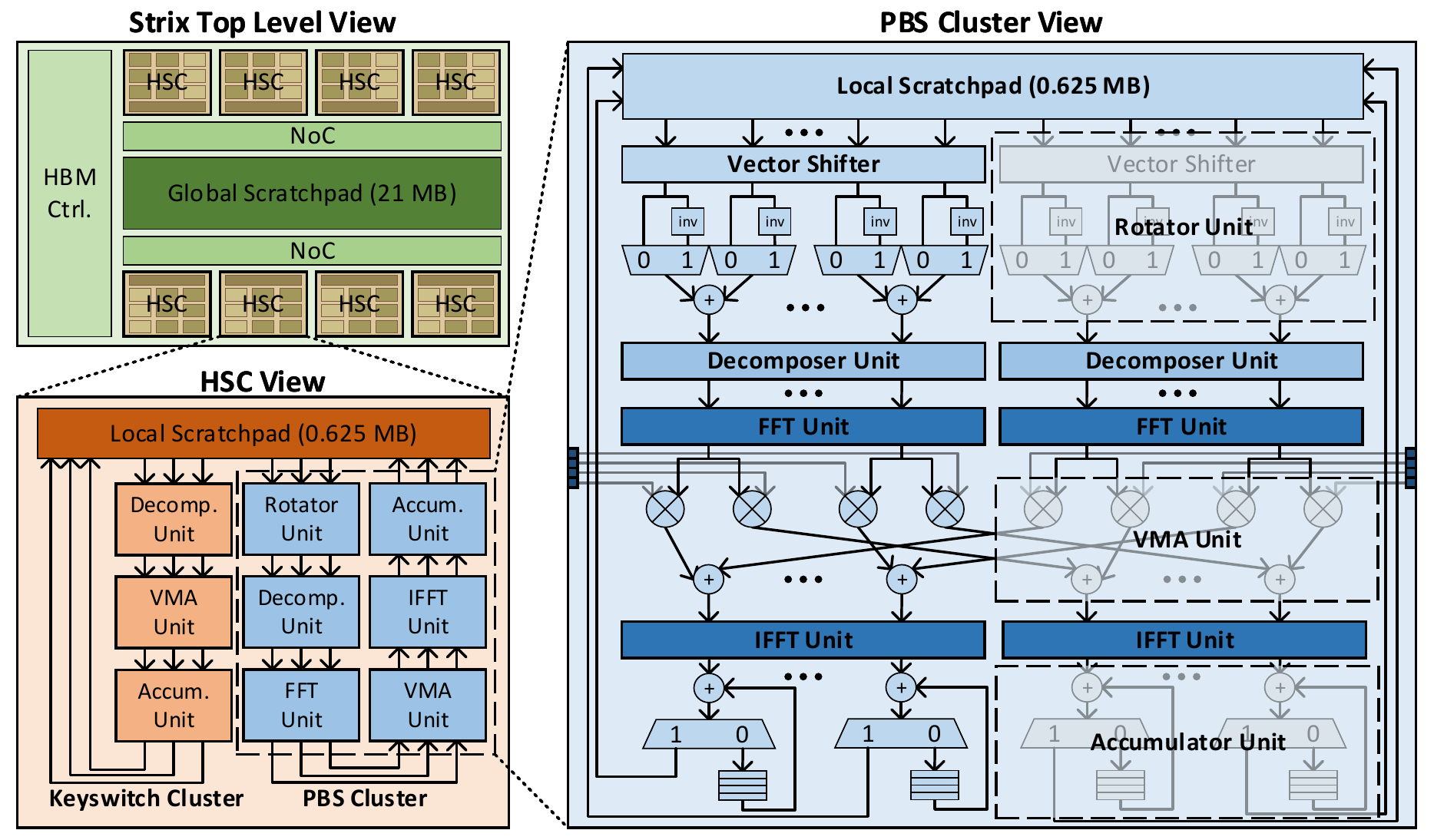}
\caption{Overall \paperN architecture and the microarchitectural detail of each functional unit inside Homomorphic Streaming Core (HSC).}
\label{fig:architecture_overall} 
\end{figure*}

% \vspace{-0.1in}
\subsection{\paperN Architecture}
Identifying four parallelisms and workload characteristics of many ciphertexts' PBS, we propose \paperN architecture optimized for stream processing of LWEs. Figure~\ref{fig:architecture_overall} depicts \paperN's overall architecture. It includes multiple Homomorphic Streaming Cores (HSC), a global scratchpad for storing keys and ciphertexts, network-on-chip (NoC) for distributing the keys to HSCs, and HBM as an external memory.

\textbf{Specialized functional units for TFHE.}
It is essential to have specialized functional units for optimizing computation and data reuse during PBS and keyswitching operations. We analyze the workload breakdown of PBS and keyswitching to identify their shared similarities. Based on our observations, we figure out that at least five distinct functional units are required to process both PBS and keyswitching. These functional units are \circlednumber{1} Rotator Unit, \circlednumber{2} Decomposer Unit, \circlednumber{3} I/FFT Unit (I/FFTU), \circlednumber{4} Vector Multiply-Add (VMA) Unit, and \circlednumber{5} Accumulator Unit.

As the name suggests, the Rotator Unit is used to perform polynomial negacyclic rotation and subtraction during blind rotation computation. The Decomposer Unit, employed in both PBS and keyswitching computations, decomposes input polynomials into several smaller value polynomials. The I/FFT Unit is utilized exclusively in blind rotation computation to transform polynomials into the frequency domain for accelerating polynomial multiplication based on the convolution theorem. The Vector Multiply-Add Unit is used in both keyswitching and PBS computations to perform element-wise multiplication between two vectors (or polynomials) and accumulate their partial sums. Finally, the Accumulator Unit aggregates all partial sum values to obtain the final sum.

\textbf{Homomorphic Streaming Core (HSC).}
All specialized functional units are instantiated within the HSC, which is responsible for executing both PBS and keyswitching operations. As depicted in Figure~\ref{fig:architecture_overall}, an HSC consists of a local scratchpad for storing test vectors and ciphertexts, along with two compute clusters: the keyswitch cluster and the PBS cluster. The PBS cluster is equipped with six-stage fully-pipelined functional units, designed to process a stream of ciphertexts in a dataflow manner without any stalling to advance the throughput of core-level batching. A full traversal of the entire pipeline in the PBS cluster corresponds to one iteration of blind rotation. We partition the computation process of a blind rotation iteration into six stages to achieve optimal workload balancing.

As discussed in Section~\ref{sec:motivation}, there is a workload imbalance issue between FFT and IFFT, because of the decomposition prior to IFFT. To address this, we split the polynomial accumulation between the frequency and time domains rather than accumulating everything in the frequency domain. This approach can yield a 1:1 workload ratio between the FFT and IFFT. Additionally, we implement a separate keyswitch cluster using a pipeline of three functional units similar to those in the PBS cluster. By incorporating separate compute clusters for both keyswitching and PBS, we can conceal the keyswitching execution time behind the PBS's and enhance the HSC's overall throughput.

\textbf{Memory system and NoC.} 
\paperN has a two-level memory structure: global scratchpad and local scratchpad. The global scratchpad is composed of several shared and private memory sections. The shared memory section of the global scratchpad stores the bootstrapping key (bsk) and keyswitching key (ksk), which are shared among all cores. To distribute the ksk and bsk among the cores, we use a fixed multicast network, as the communication is one-to-all and unidirectional. The private memory section of the global scratchpad stores the LWE ciphertexts and the initial test vector inputs for each core. Since the private memory section is exclusive to each core, we utilize a point-to-point network for reading and writing data between the private memory sections and the cores.

Conversely, the local scratchpad, located within each core, stores intermediate test vectors and LWE ciphertexts. Similar to the global scratchpad, the local scratchpad contains separate sections allocated for the PBS cluster and keyswitch cluster memory. The PBS cluster memory is used exclusively for storing intermediate test vectors generated during blind rotation computations. In the final iteration of blind rotation, the output LWE ciphertext is stored directly in the keyswitch cluster memory, initiating the keyswitching operation. As depicted in Figure~\ref{fig:architecture_overall}, the local scratchpad can only be accessed by specific functional units, and since the communication pattern between functional units and the local scratchpad is fixed, a point-to-point network can be employed.

\subsection{Workload Scheduling}
\label{algorithmic pipelines} 
In any privacy-preserving computation using TFHE, the workload can be regarded as a series of sequential PBS and keyswitching operations. Taking the neural network example, the network is first transformed into a series of linear and nonlinear operations. Linear operations, such as homomorphic addition and scalar multiplication on ciphertexts, can be executed rapidly compared to nonlinear operations. Nonlinear operations, which involve evaluating the activation function, are necessary for both PBS and keyswitching.

\paperN schedules the workload in a series of epochs, with each epoch containing a maximum number of LWEs equal to the product of device-level and core-level batch sizes. The device-level batch size is determined by the number of compute cores (HSCs), while the core-level batch size depends on the number of LWE test-vectors that can be stored in the local scratchpad, which varies based on the chosen TFHE parameters. The PBS cluster within each core executes $n$ iterations of blind rotation, processing a core-level batch size of LWEs in a pipeline through all functional units in the PBS cluster during each iteration.

Once the blind rotation is complete, the output is forwarded to the keyswitching cluster to begin the keyswitching process, while simultaneously initiating a new blind rotation for the next epoch to hide the latency of the keyswitching process. The keyswitching operation is typically faster than the blind rotation, as it involves standard matrix-matrix multiplication without any polynomial operations. The keyswitch cluster computes the matrix-matrix multiplication by tiling it into a series of smaller matrix-matrix multiplications and working on each smaller matrix as a sequence of vector-matrix multiplications. During these processes, both the bootstrapping key and keyswitching key are distributed from the global scratchpad, while the next portions of the bootstrapping and keyswitching keys for the subsequent iteration and tile are fetched from external memory. After all operations are complete, the output is collected in the global scratchpad and written back to the host machine.

\section{Microarchitecture}
\label{sec:micro}

%We have implemented \paperN based on the parallelism identified in Section~\ref{architecting}. Specifically, we have chosen $TvLP=8$, $CLP=4$, $PLP=2$, and $CoLP=2$.
%To achieve high throughput, we have designed HSC as a streaming processor with various specialized functional units, where each unit takes in a stream of data and produces a stream of data that is consumed by the next unit. This design is similar to the vector chaining technique used in vector processors\cite{russell1978cray}, providing the advantage of simplifying memory access on the local scratchpad.

\begin{figure}[t]
    \centering
    \includegraphics[width=\linewidth]{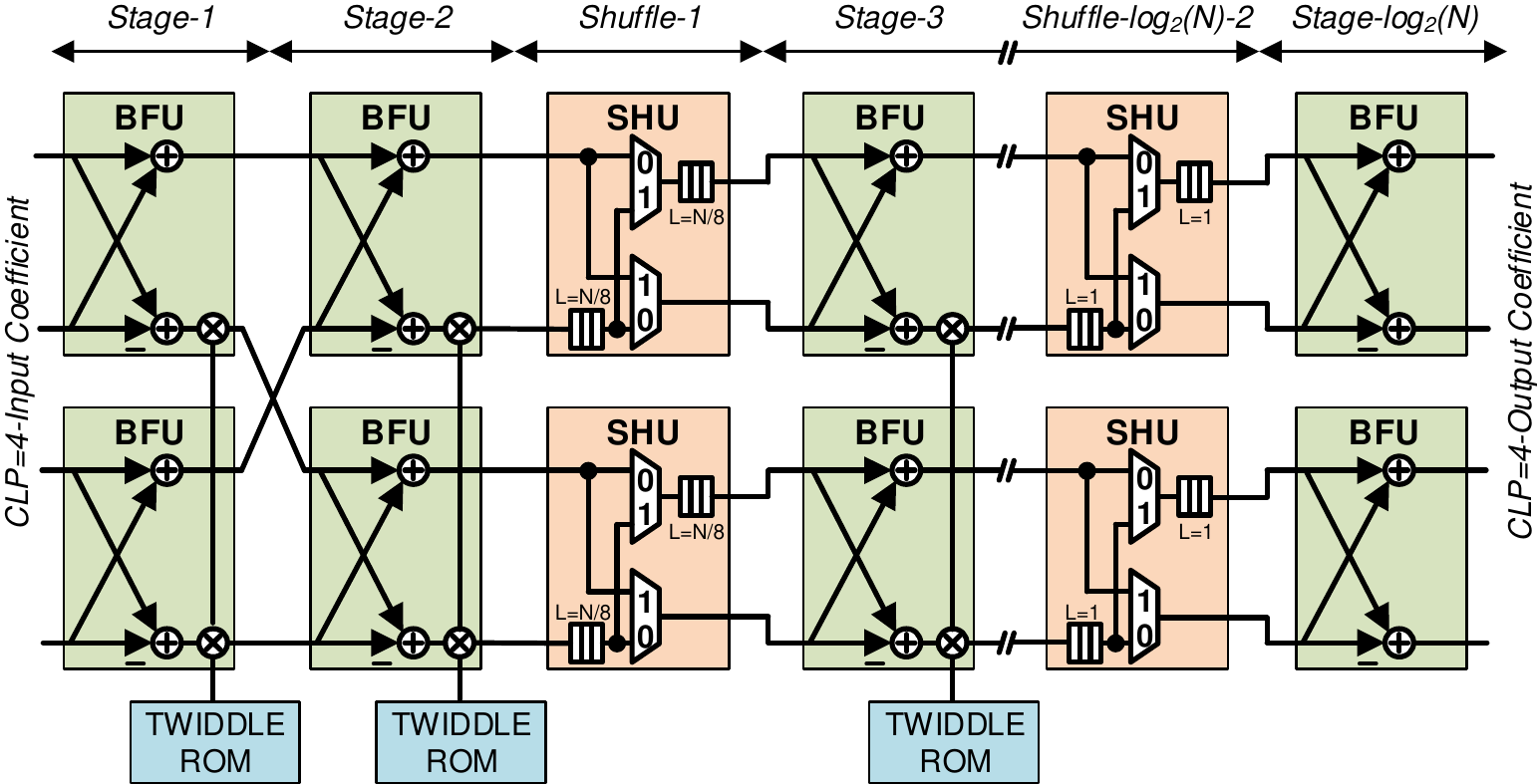}
    % \vspace{-0.3in}
    \caption{N-point of pipelined-(I)FFT unit}
    \label{fig:FFT}
\end{figure}

\subsection{I/FFT Unit}
\paperN utilizes the FFT algorithm to compute the polynomial multiplication.
It is well-known that the FFT algorithm consists of many similar computations called butterfly, which operates on two complex-number inputs and the twiddle factor of FFT. Even though this offers the opportunity for parallelism, the required input and the connection between the butterfly are not straightforward. In general, the FFT algorithm heavily relies on data shuffling to match up the connection between the butterfly. As the number of points of FFT increases, it is difficult to do all the computations simultaneously. It is required to store the partial results and reload them when they are needed again for the butterfly operation. However, it causes an irregular memory access pattern due to the characteristics of the butterfly connection in FFT. To solve this problem, we design the pipelined-FFT, which has the computation through all the stages in a pipeline manner. With this architecture, the data shuffle can be done inside the (I)FFT unit because the partial result is directly needed for the next stage butterfly.
% The FFT consists of numerous butterfly computations, which enable parallelism but involve complex input connections and data shuffling. As the number of FFT points increases, computations become challenging due to irregular memory access patterns. To address this, we designed a pipelined-FFT architecture, allowing data shuffling within the (I)FFT unit and streamlining butterfly computations.

Figure~\ref{fig:FFT} shows the FFT/IFFT unit based on the pipelined-FFT architecture \cite{garrido2021survey}. This unit exploits the four-lane input coefficient of polynomial corresponding to the $CLP=4$ parallelism. It consists of $log_2 (N)$ stage of butterfly units (BFUs), which associate with all the stages computation in $N$-point FFT. The number of butterflies in each stage depends on the number of $CLP$. In this case ($CLP=4$), we need 2 BFUs per stage. All the data-path in the unit consist of two fixed-point numbers for the real and imaginary parts. The shuffle unit (SHU) connects the different stages of the butterfly. It shuffles the data depending on the number of delays, i.e., $L$, in each stage. It has two $L$-delay elements and two multiplexers to alternate the data stream from the upper input and bottom input. The multiplexer selector signal is commuted every $L$ clock cycles to shuffle the input stream. In the implementation, we use a shift register for the small delay element $(L < 32)$ and use an SRAM-based shift register for the large delay element $(L >= 32)$ for the efficiency of the area and energy. We have the Twiddle-ROM at every stage to accommodate that every BFU needs different twiddle factor. Although this architecture requires additional memory space, it achieves high throughput. The initial latency for transforming an $N-1$ degree polynomial is $\frac{N}{CLP}$. Afterward, it can transform an $N-1$ degree polynomial every $\frac{N}{CLP}$ clock cycles consecutively, maintaining maximum throughput.

\textbf{Folding optimization.}
In general, every signal, including polynomials, can be decomposed into four components: real-even, imaginary-even, real-odd, and imaginary-odd signals. FFT has symmetry properties that can map these components cleanly to the frequency domain\cite{rabiner1979use,guo1998quick}. When performing polynomial transformations, these symmetry properties can be used to combine two polynomials in a single FFT operation, placing one polynomial in the real part and another in the imaginary part. This optimization method is used in the latest TFHE library\cite{chillotti2020concrete}. Although this mapping appears efficient because it conserves hardware resources by transforming two polynomials with a single FFT unit, an additional hardware module is needed after the Fourier transform to separate the first and second signals. This extra module requires memory buffers to reorder the signals, significantly increasing overall latency and area.

A more effective way to utilize the imaginary slot in the Fourier transform is to employ the folding scheme\cite{klemsa2021fast}. This scheme splits an $N-1$ degree polynomial into two $\frac{N-1}{2}$ degree polynomials, placing one polynomial as the imaginary component of the other, effectively folding and reducing the signal length by half. Although this means we need as many FFT modules as there are polynomials to transform in parallel, it eliminates the additional hardware overhead for signal separation. Moreover, by halving the signal length, we can implement smaller FFT modules in the hardware.
\begin{figure}[t]
    \centering
    \includegraphics[width=\linewidth]{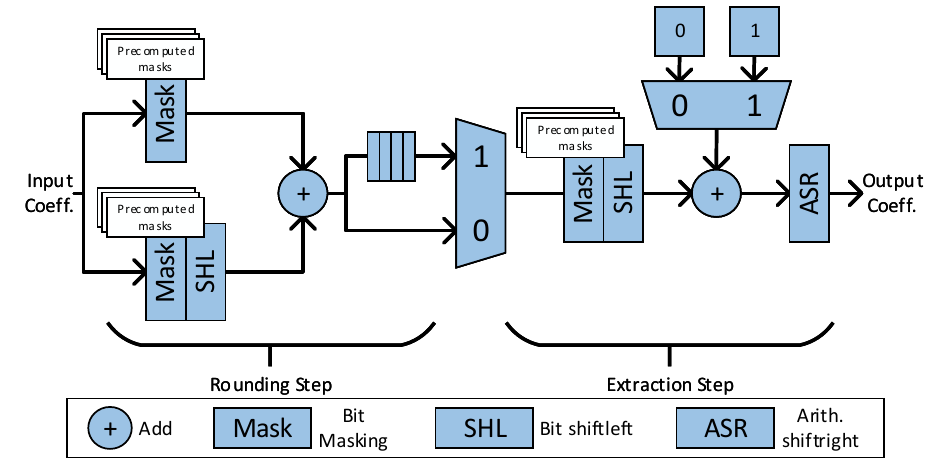}
    \caption{Microarchitectural details of a single lane within the fully pipelined decomposer unit.}
    \label{fig:decomposer}
\end{figure}
The folding scheme allows our FFT units to virtually support $CLP=8$ by only implementing $CLP=4$, while also reducing the size of the FFT unit. In our case, the FFT unit was initially designed to support up to $16,384$-point polynomial transformations, but we reduced it to $8,192$-point transformations by applying the folding scheme.
Although we must implement $CLP=8$ for other functional units besides the FFT unit to fully benefit from the folding scheme, increasing $CLP$ for other functional units is less expensive compared to increasing $CLP$ in the FFT unit. Consequently, the folding scheme substantially reduces the size of our FFT unit by almost 50\%.

\subsection{Decomposer Unit}
The decomposer unit receives a polynomial as input and decomposes it into $l_b$ polynomials. However, this decomposition differs from the commonly used RNS decomposition in CKKS accelerators. The decomposition process for each coefficient in the polynomial is carried out according to the following expression:
\begin{equation}
    \left\lVert a - \sum_{i=1}^{l_b} \left\lfloor \frac{a}{Q/{B^i}} \right\rfloor_B \right\rVert_{\infty} \leq \frac{Q}{B^{l_b}}
\end{equation}
where $a$ is the polynomial's coefficient, $Q$ is the original ciphertext modulo, $B$ is the new basis for decomposition, and $l_b$ is the decomposition level.

The challenge lies in designing a fully pipelined decomposer unit that can accept input and produce output every cycle, thereby maintaining maximum throughput. Additionally, the decomposition process must be executed for each coefficient in the polynomial. As evident, the required division and rounding operations are quite complex and expensive. To address this issue, we propose a novel decomposer unit that divides the process into two steps: rounding and extraction, as illustrated in Figure~\ref{fig:decomposer}. In order to mitigate the high costs, we implement the entire circuit without using any multipliers. By employing a combination of masking, shifting, and adding, we reduce both the circuit cost and overall complexity.

In the rounding step, the coefficient term is rounded to the nearest representable value according to the chosen TFHE parameter. We use a combination of masks to extract the required upper bits and carry to obtain the rounded coefficient. This rounded coefficient is then decomposed into $l_b$ coefficients across multiple clock cycles, and we store them in a buffer for later use. In the extraction step, we utilize a series of prepared masks to extract specific bit portions. After obtaining the extracted bit, we add it to the carry (zero or one) from the previous extracted bit to derive the final decomposed value. The decomposer unit operates for $\frac{N}{CLP} \times l_b$ cycles for each polynomial. To align the throughput with the rotator and FFT units, we implement the decomposer unit with $2 \times CLP$ lanes, achieving a chip-wide throughput of $2 \times CLP \times CoLP \times TvLP$ data per clock cycle.

\subsection{Rotator, VMA, and Accumulator Unit}

\textbf{Rotator unit.}
The rotator unit, located within the PBS cluster, is responsible for carrying out negacyclic rotation and polynomial subtraction as in Algorithm \ref{algorithm_pbs}. As depicted in Figure~\ref{fig:architecture_overall}, the rotator unit's datapath employs eight lanes ($2 \times CLP$) for processing eight coefficients in parallel, aligning with the folding scheme's requirements. In each clock cycle, the rotator unit reads two coefficients from each memory bank for both polynomials, performs cyclic rotation to align the coefficients with the lanes, and adds/subtracts polynomials. 
With $CoLP=2$, we utilize two instances of the rotator unit, achieving a chip-wide throughput of $2\times CLP \times CoLP \times TvLP$ coefficients per clock cycle and one initiation interval.

\textbf{VMA unit.}
The VMA unit is designed to compute the vector-matrix multiplication of polynomials between test-vector and bsk polynomials. Coefficients of the bsk polynomial are sourced from the global scratchpad and distributed via the multi-cast network. The VMA units in the PBS cluster and keyswitching cluster have different data types, with the former utilizing complex multipliers and adders and the latter employing integer multipliers and adders. Figure~\ref{fig:architecture_overall} illustrates the VMA unit's architecture, which includes multipliers followed by an adder tree that accumulates the results into a partial sum. With $PLP=2$, two instances of the VMA unit are utilized, yielding a chip-wide throughput of $CLP \times PLP \times TvLP$ complex coefficients per clock cycle.

\textbf{Accumulator unit.}
Once the VMA unit has accumulated data via an adder tree, the output is directed to the IFFT unit for transformation back into the time domain as a polynomial. The IFFT unit's output is then sent to the accumulator unit for final accumulation in the time domain. The accumulator unit consists of $2 \times CLP$ input lanes, each containing an accumulator with a buffer (see Figure~\ref{fig:architecture_overall}). Each buffer can store $\frac{N}{2 \times CLP}$ coefficients. When the final sums are obtained, they are forwarded back to the local scratchpad for use as input in the next blind rotation iteration. By utilizing two instances of the accumulator unit with $CoLP=2$, the overall throughput rate is $2 \times CLP \times CoLP \times TvLP$ coefficients per clock cycle.

\begin{table}[t]
\centering
\caption{The area and power breakdown of \paperN.}
\label{table:formatting}
\normalsize
\sisetup{table-format=3.2}
\begin{tabular}{lS[table-format=2.2]S}
\toprule
Component & {\begin{tabular}[c]{@{}l@{}}Area\\ (\si{mm^2})\end{tabular}} & {\begin{tabular}[c]{@{}l@{}}Power\\ (\si{W})\end{tabular}} \\
\midrule
Local scratchpad (0.625MB) & 0.92 & 0.47 \\
Rotator & 0.02 & 0.01 \\
Decomposer & 0.28 & 0.02 \\
I/FFTU & 7.23 & 5.49 \\
VMA & 0.63 & 0.10 \\
Accumulator & 0.32 & 0.13 \\
\midrule
\textbf{1 core} & 9.38 & 6.21 \\
\midrule
8 cores & 75.03 & 49.67 \\
Global NoC & 0.04 & 0.01 \\
Global scratchpad (21MB) & 51.40 & 26.24 \\
HBM2 PHY & 14.90 & 1.23 \\
\midrule
\textbf{Total} & 141.37 & 77.14 \\
\bottomrule
\end{tabular}
\end{table}
\section{Evaluation}
\subsection{Hardware Modeling}

We implement \paperN's functional units, scratchpads, and interconnections in System Veriog and synthesize them using the TSMC 28nm process design kit. All functional unit datapaths are 32-bit, except for the FFTU which uses 64-bit and twiddle 16-bit to ensure accuracy. We have then further optimized our design to minimize the area and power consumption while maintaining 1.2 GHz clock frequency. We construct the global scratchpad with a total capacity of $21$ MB as a dual-port (one read, one write), multi-bank 128-bit width SRAM.
It is also implemented as double-buffered to store bsk, ksk, and (R)LWE ciphertext during computation to decouple off-chip memory access and the computation. Assuming a modest bandwidth of 300 GB/s for one HBM2e stack\cite{jedecHBM}, we utilize 8 channels for bsk's transfer, 4 channels for ksk's transfer, and the remaining 4 channels for transferring ciphertexts. 
Moreover, we implement a scratchpad inside each HSC as a true dual-port multi-bank 32-bit SRAM with $0.625$ MB capacity. We implement two separate data buses to broadcast bsk and ksk, where each bus has 512-bit and 256-bit widths, respectively. Table~\ref{table:formatting} shows the synthesis result of the \paperN with 8 HSCs. Overall, it takes $141.37$ mm$^2$ area and consumes up to $77.14$ $W$ power.

\begin{table}[t]
  \centering
  \caption{TFHE parameter set for the experiments.}
  \small
  \begin{tabular}{@{}crrrr}
    \toprule
    \multirow{2}{*}{} & {I} & {II} & {III} & {IV} \\ 
    \cmidrule(lr){2-5}
    $n$ & 500 & 630 & 592 & 991 \\ 
    $k$ & 1 & 1 & 1 & 1  \\
    $N$ & 1024 & 1024 & 2048 & 16384 \\ 
    $l_\mathrm{b}$ & 2 & 3 & 3 & 2 \\
    $\lambda$ & 110-bit & 128-bit & 128-bit & 128-bit \\
    \bottomrule
  \end{tabular}
  \label{tab:security_param}
\end{table}

\begin{table}[t]
  \centering
  \caption{Comparison of PBS latency and throughput on a variety of platforms}
  \normalsize
  \resizebox{0.9\columnwidth}{!}{ 
    \begin{tabular}{@{}lllS[table-format=4.2]S[table-format=5.0,group-separator={,},group-minimum-digits=4]@{}}
      \toprule
      & Platform & Parameter Set & {Latency} & {Throughput} \\ 
      & & & {(\si{\milli\second})} & {(\si{PBS\per\second})} \\
      \midrule
      \multirow{4}{*}{Concrete\cite{chillotti2020concrete}} & \multirow{4}{*}{CPU} & I & 14.00 & 70 \\
      & & II & 19.00 & 52 \\
      & & III & 38.00 & 26 \\
      & & IV & 969.00 & 1 \\
      \cmidrule(lr){2-5}
      NuFHE\cite{nufhe} & \multirow{2}{*}{GPU} & I & 37.00 & 2000 \\
      & & II & 700.00 & 500 \\
      \cmidrule(lr){2-5}
      \multirow{2}{*}{YKP\cite{ye2022HPEC}} & \multirow{2}{*}{FPGA} & I & 1.88 & 2657 \\
      & & III & 4.78 & 836 \\
      \cmidrule(lr){2-5}
      \multirow{2}{*}{XHEC\cite{nam22nbit}} & & I & {--} & 2200 \\
      & \multirow{-2}{*}{FPGA} & II & {--} & 1800 \\
      \cmidrule(lr){2-5}
      Matcha\cite{jiang2022matcha} & ASIC & I & 0.20 & 10000 \\
      \cmidrule(lr){2-5}
      \multirow{4}{*}{\paperN} & \multirow{4}{*}{ASIC} & I & 0.16 & 74696 \\
      & & II & 0.23 & 39600 \\
      & & III & 0.44 & 21104 \\
      & & IV & 3.31 & 2368 \\
      \bottomrule
    \end{tabular}
  }
  \label{tab:microbenchmark}
\end{table}

\subsection{Experimental Methodology}
To test the performance of \paperN, we create a custom cycle-level simulator and model the accelerator in an object-oriented manner to determine the latency, bandwidth, and throughput of each functional unit. Specifically, the simulator converts the input workload as a computational graph with nodes, where each node mainly represents either bootstrapping or keyswitching or a combination of both operations. Each node in the graph will further be decomposed into several blind rotation fragments.

To evaluate the performance of \paperN, we conduct microbenchmark and application benchmark tests. The microbenchmark assesses the latency and throughput of PBS and keyswitching operations. The simulator allows us to set the desired number of LWE ciphertexts and various TFHE parameters to compute the required number of blind rotate fragments. We compared the results of \paperN with previous works, including the original Concrete library (Intel Xeon Platinum)\cite{chillotti2020tfhe}, nuFHE (Nvidia Titan RTX)\cite{nufhe}, YKP (FPGA)\cite{ye2022HPEC}, XHEC (FPGA)\cite{nam22nbit}, and Matcha(ASIC)\cite{jiang2022matcha}. 

Table~\ref{tab:security_param} specifies the TFHE parameter sets we used for the experiments. Parameter set \rmfamily{I} serves as a baseline comparison, which has been used in all prior works. Set \rmfamily{II} and \rmfamily{III} are specific case parameters used by YKP and XHEC, respectively. Meanwhile, set \rmfamily{IV} is new TFHE parameters tested on \paperN, with the highest polynomial degree corresponding to a heavier computational workload and better precision. We also perform experiments to evaluate the effects of FFT optimization on our design, using parameter set \rmfamily{I}. Additionally, we showcase our timing measurements for functional units to demonstrate their utilization and bandwidth efficiency. For the application benchmark, we conducted tests on Zama Deep-NN models with 20, 50, and 100-layer depths, using the same TFHE parameters and polynomial degrees ($N=1024$, $N=2048$, and $N=4096$) as in the reference paper \cite{chillotti2021programmable}. 

\begin{figure}[t]
    \centering
    \includegraphics[width=\linewidth]{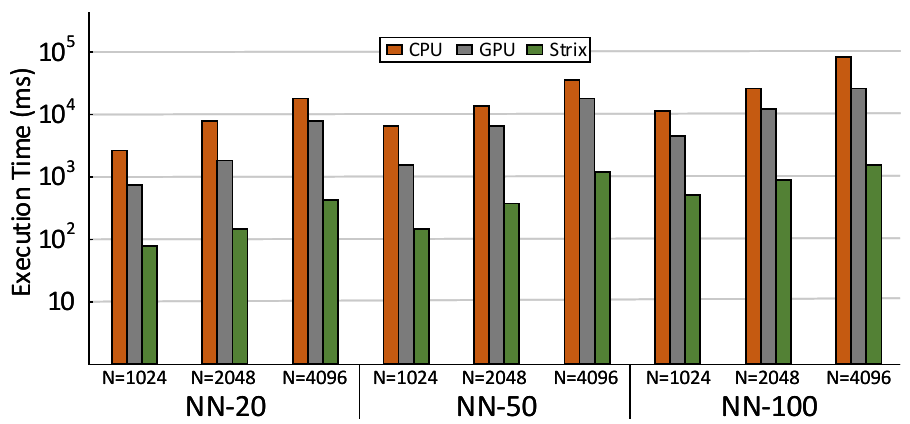}
    \caption{Application benchmark results using Zama Deep-NN models~\cite{chillotti2021programmable}. The neural network execution time was compared against CPU and GPU.}
    \label{fig:NN_benchmark}
\end{figure}

\subsection{Results and Discussions}

\textbf{Throughput and latency.} Table~\ref{tab:microbenchmark} displays the latency and throughput performance of \paperN in comparison to previous works. As observed, \paperN achieves superior performance in terms of throughput and latency compared to all prior works. Additionally, \paperN supports a wider range of TFHE parameter sets than any other previous accelerator. When compared to the previous state of the art, Matcha, which only supports parameter set I, \paperN attains a 7.4$\times$ improvement in throughput and a 1.25$\times$ reduction in latency. The modest latency enhancement is expected since Strix only utilizes a modest amount of $CLP=4$. Conversely, the GPU implementation, NuFHE, solely supports TFHE parameters with a fixed polynomial degree $N=1024$ and is optimized only for $l_b=2$. Changing the parameter set from \rmfamily{I} to \rmfamily{II} in NuFHE significantly reduces performance, as they discard the blind rotation kernel and execute the entire blind rotation using the FFT kernel sequentially. Lastly, \paperN exhibits its exceptional performance for the largest TFHE parameter set \rmfamily{IV}, improving the throughput and latency by 2,368$\times$ and 292$\times$, respectively, in comparison to Concrete.

\textbf{Zama Deep-NN.}
Zama Deep-NN has three different models: NN-20, NN-50, and NN-100, which indicate the depth of the neural network. \textcolor{black}{The input consists of $28\times28$ pixels, where each pixel is encrypted with one cipher. The first layer performs $10 \times 11$ convolution followed by ReLU activation, producing an output image of dimensions [1,2,21,20]. The remaining layers are dense layers with 92 neurons on each layer, followed by ReLU activation between each layer. To perform every ReLU operation, we utilize PBS}. We conduct three different TFHE parameter setups for each model, increasing the polynomial size and the security level, which increases the workload. The parameters for the experiment are detailed in \cite{chillotti2021programmable}. Figure~\ref{fig:NN_benchmark} shows the execution time results of running Zama Deep-NN on \paperN. In all cases, \paperN outperforms the CPU and GPU implementations. We observe that \paperN speeds up the evaluation time for NN by $33-38\times$ compared to CPU and $8-17\times$ compared to GPU. Furthermore, we note that \paperN's speedup becomes more evident with heavier workloads compared to CPU and GPU.

\begin{table}[t]
\centering
\caption{FFT optimization effects on performance.}
\label{table:fft_effect}
\small
\begin{tabularx}{\columnwidth}{Xcccc}
\toprule
\textbf{Metric} & \textbf{No Fold.} & \textbf{With Fold.} & \textbf{Improv.} \\
\midrule
Latency (ms) & 0.27 & 0.16 & 1.68$\times$ \\
Throughput (PBS/s) & 37,472 & 74,696 & 1.99$\times$ \\
FFT Unit Area (mm$^2$) & 3.13 & 1.81 & 1.73$\times$ \\
Total Core Area (mm$^2$) & 13.87 & 9.38 & 1.48$\times$ \\
\bottomrule
\end{tabularx}
%\vspace{-0.1in}
\end{table}

\textbf{FFT optimization impact on area and performance.}
Unlike leveled HE schemes that utilize NTT, TFHE can employ FFT for polynomial multiplication, presenting new opportunities for algorithm-hardware optimization. We implement a folding scheme in our FFT algorithm, as explained in Section~\ref{sec:micro}. To assess the impact of the folding scheme on overall performance and design, we create two types of \paperN: non-folded \paperN and folded \paperN, both with $CLP=4$ for the $16,384$-point FFT unit. Table~\ref{table:fft_effect} compares non-folded \paperN and folded \paperN in terms of latency, throughput, FFT unit area, and total core area, which have been improved by 1.68$\times$, 1.99$\times$, 1.73$\times$, and $1.48\times$, respectively. Without folding optimization, all units in the core must utilize a 4-lane implementation to match the bandwidth of the 4-lane FFT unit, affecting the final latency and throughput performance. Additionally, a $16,384$-point FFT unit would cost twice more hardware resources than our current design.

\begin{figure}[t]
    \centering
    \includegraphics[width=\linewidth]{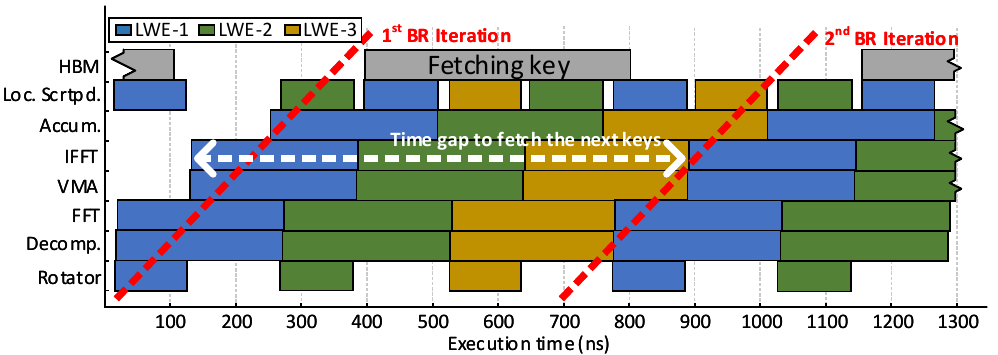}
    \caption{\textcolor{black}{The experiment measures the timing of all functional units for the first two iterations of blind rotation, assuming that each core processes three LWE ciphertexts in each iteration.}}
    \label{fig:timing}
\end{figure}

\textbf{Overall timing and bandwidth utilization.}
We conducted measurements to analyze how \paperN's hardware and bandwidth are being utilized. Figure~\ref{fig:timing} shows the overall timing for the first two iterations of blind rotation within each core, assuming that each core processed three LWE ciphertexts (represented by different colors) using parameter set I. All functional units in \paperN are highly utilized, with the decomposer, I/FFT, VMA, and accumulator units achieving close-to 100\% utilization rates, while the rotator unit reached a 50\% utilization rate. The local scratchpad was also in high demand, being accessed about 90\% of the time by the rotator unit for read accesses and the accumulator unit for write accesses. Moreover, the HBM bandwidth was occupied around 60\% of the time, primarily used for transfering the bootstrapping key during blind rotation. For each iteration, the bootstrapping key needed to be fetched only once and was then reused for all LWEs. The HBM bandwidth utilization was significantly influenced by the number of LWEs processed for each iteration and the number of lanes ($CLP$). If the number of LWEs is small or $CLP$ is high, the system would be bottlenecked by HBM (i.e., memory-bound) as the time gap to fetch the next key became smaller. Conversely, if the number of LWEs is large or $CLP$ is low, the system would be bottlenecked by the (I)FFT units (i.e., compute-bound).

\textbf{TvLP vs CLP trade off.}
We instantiate several versions of \paperN with a variety of $TvLP$ and $CLP$ values to examine how they affect throughput and latency performance when the maximum external bandwidth is fixed to 300 GB/s, and to determine the required bandwidth for each particular $TvLP$ and $CLP$ configuration. We perform this experiment using parameter set IV. Table~\ref{tab:tradeoff} illustrates how a variety of $TvLP$ and $CLP$ values impact performance while keeping the product of $TvLP$ and $CLP$ constant. When $TvLP=16$ and $CLP=2$, \paperN becomes heavily compute-bound, requiring only 200 GB/s bandwidth and yielding a latency performance of 7.2 ms. We observe that the $TvLP=8$ and $CLP=4$ configuration represents the sweet spot, balancing the compute and memory requirements. It requires a bandwidth of 257 GB/s, achieving 3.8ms latency while sustaining high throughput of 2368 PBS/s. The remaining $TvLP$ and $CLP$ configurations with $CLP$ values greater than 4 render the system memory-bound, demanding bandwidth higher than 300 GB/s and causing underutilization of the compute core. In the extreme case of instantiating only one core ($TvLP = 1$ and $CLP=32$), the required bandwidth is about 1 TB/s, making the workload heavily memory-bound and resulting in a throughput drop to only 620 PBS/s. As in our \paperN implementation, the configuration of $TvLP=8$ and $CLP=4$ provides the best performance given the 300 GB/s external memory bandwidth.

\begin{table}[t]
  \centering
  \caption{The TvLP and CLP effects on throughput, latency, and the required external bandwidth}
  \normalsize
  \resizebox{\columnwidth}{!}{
    \begin{tabular}{@{}lS[table-format=2]S[table-format=4,group-separator={,},group-minimum-digits=4]S[table-format=1.1,group-separator={,},group-minimum-digits=4]S[table-format=4,group-separator={,},group-minimum-digits=4]@{}}
      \toprule
      {TvLP} & {CLP} & {\multirow{2}{*}{\begin{tabular}[c]{@{}l@{}}Throughput\\ (PBS/s)\end{tabular}}} & {\multirow{2}{*}{\begin{tabular}[c]{@{}l@{}}Latency\\ (ms)\end{tabular}}} & {\multirow{2}{*}{\begin{tabular}[c]{@{}l@{}}Required Bandwidth\\ (GB/s)\end{tabular}}} \\
             &       &                                                                                       &                                                 &                                                                                                   \\
      \midrule
      16     & 2     & 2368                                                                                   & 7.2                                               & 200                                                                                             \\
      8      & 4     & 2368                                                                                   & 3.8                                               & 257                                                                                             \\
      4      & 8     & 2364                                                                                   & 3.8                                               & 371                                                                                             \\
      2      & 16    & 1240                                                                                   & 3.6                                               & 599                                                                                             \\
      1      & 32    & 620                                                                                    & 3.6                                               & 1053                                                                                            \\
      \bottomrule
    \end{tabular}
  }
  \label{tab:tradeoff}
\end{table}
\section{Related Works}

\textbf{CKKS-based accelerator.}
Recent research on FHE accelerators primarily focuses on accelerating CKKS schemes with designs like FAB\cite{agrawal2022fab}, CraterLake\cite{samardzic2022craterlake}, and ARK\cite{kim2022ark}. Although direct comparison is difficult, CKKS accelerators likely perform better on large machine learning workloads due to inherent packing operations. However, TFHE supports more applications with its universal operations and can be designed with lower hardware costs, particularly in memory footprint. Our design requires 26 MB of on-chip memory, compared to FAB, CraterLake, and ARK's 45 MB, 256 MB, and 512 MB, respectively. TFHE's working set can be less than 1 MB, allowing further memory reduction under resource constraints. CKKS acceleration demands higher external memory bandwidth as it is memory-bound\cite{de2021does}, while TFHE is compute-bound. Most CKKS accelerators need two stacks of HBM2e for 1 TB/s bandwidth, while our design requires only one stack of HBM2e 300 GB/s.

\textbf{TFHE-based accelerator}
Various works have attempted to accelerate the TFHE scheme on platforms like GPU \cite{nufhe, dai2015cuhe, chillotti2020concrete}, FPGA \cite{gener2021fpga,ye2022HPEC, nam22nbit}, and ASIC \cite{jiang2022matcha}. NuFHE's GPU acceleration uses either NTT or FFT, with FFT outperforming NTT \cite{nufhe}. In FPGA acceleration, \cite{ye2022HPEC} employs NTT for polynomial multiplication, achieving a 1.3$\times$ throughput improvement compared to NuFHE. ASIC acceleration in Matcha \cite{jiang2022matcha} outperforms prior accelerators using bootstrapping key unrolling \cite{bourse2018fast} that reduces blind rotation iterations at the cost of increased key size. This paper presents \paperN, which advances TFHE acceleration by utilizing two-level batching and specialized functional units. Our streaming architecture-based accelerator supports various TFHE parameters and achieves 7.4$\times$ higher throughput than Matcha.

% \vspace{-0.12in}
\section{Conclusion}
\paperN is the first accelerator specifically designed to enable high-throughput PBS for the TFHE scheme. We first identify the challenges of accelerating TFHE on GPUs and recognize the blind rotation fragmentation problem due to limited batching capabilities. To address this issue, we propose device-level batching and core-level batching to significantly increase the batch size. However, to efficiently implement core-level batching, specialized and optimized functional units are required. To achieve this, we analyze the parallelism levels in PBS and identify four levels of parallelism that can be exploited. Based on these insights, we design specialized, fully-pipelined functional units that enable streaming ciphertext processing for high-throughput performance. \paperN has achieved $1,067\times$ and $37\times$ throughput improvements compared to CPUs and GPUs, respectively, and attained $38\times$ and $17\times$ latency improvements when evaluating deep neural networks compared to CPUs and GPUs, respectively.

% \section*{Acknowledgment}

% The preferred spelling of the word ``acknowledgment'' in America is without 
% an ``e'' after the ``g''. Avoid the stilted expression ``one of us (R. B. 
% G.) thanks $\ldots$''. Instead, try ``R. B. G. thanks$\ldots$''. Put sponsor 
% acknowledgments in the unnumbered footnote on the first page.

%%%%%%%%% -- BIB STYLE AND FILE -- %%%%%%%%
\bibliographystyle{IEEEtran}
\bibliography{refs}

\begin{thebibliography}{10}
\providecommand{\url}[1]{#1}
\csname url@samestyle\endcsname
\providecommand{\newblock}{\relax}
\providecommand{\bibinfo}[2]{#2}
\providecommand{\BIBentrySTDinterwordspacing}{\spaceskip=0pt\relax}
\providecommand{\BIBentryALTinterwordstretchfactor}{4}
\providecommand{\BIBentryALTinterwordspacing}{\spaceskip=\fontdimen2\font plus
\BIBentryALTinterwordstretchfactor\fontdimen3\font minus
  \fontdimen4\font\relax}
\providecommand{\BIBforeignlanguage}[2]{{%
\expandafter\ifx\csname l@#1\endcsname\relax
\typeout{** WARNING: IEEEtran.bst: No hyphenation pattern has been}%
\typeout{** loaded for the language `#1'. Using the pattern for}%
\typeout{** the default language instead.}%
\else
\language=\csname l@#1\endcsname
\fi
#2}}
\providecommand{\BIBdecl}{\relax}
\BIBdecl

\bibitem{rivest1978data}
R.~L. Rivest, L.~Adleman, M.~L. Dertouzos \emph{et~al.}, ``On data banks and
  privacy homomorphisms,'' \emph{Foundations of secure computation}, vol.~4,
  no.~11, pp. 169--180, 1978.

\bibitem{gentry2009fully}
C.~Gentry, ``Fully homomorphic encryption using ideal lattices,'' in
  \emph{Proceedings of the forty-first annual ACM symposium on Theory of
  computing}, 2009, pp. 169--178.

\bibitem{lee2022privacy}
J.-W. Lee, H.~Kang, Y.~Lee, W.~Choi, J.~Eom, M.~Deryabin, E.~Lee, J.~Lee,
  D.~Yoo, Y.-S. Kim \emph{et~al.}, ``Privacy-preserving machine learning with
  fully homomorphic encryption for deep neural network,'' \emph{IEEE Access},
  vol.~10, pp. 30\,039--30\,054, 2022.

\bibitem{jung2021over}
W.~Jung, S.~Kim, J.~H. Ahn, J.~H. Cheon, and Y.~Lee, ``Over 100x faster
  bootstrapping in fully homomorphic encryption through memory-centric
  optimization with gpus,'' \emph{IACR Transactions on Cryptographic Hardware
  and Embedded Systems}, pp. 114--148, 2021.

\bibitem{shivdikar2022accelerating}
K.~Shivdikar, G.~Jonatan, E.~Mora, N.~Livesay, R.~Agrawal, A.~Joshi, J.~L.
  Abell{\'a}n, J.~Kim, and D.~Kaeli, ``Accelerating polynomial multiplication
  for homomorphic encryption on gpus,'' in \emph{2022 IEEE International
  Symposium on Secure and Private Execution Environment Design (SEED)}.\hskip
  1em plus 0.5em minus 0.4em\relax IEEE, 2022, pp. 61--72.

\bibitem{dai2015cuhe}
W.~Dai and B.~Sunar, ``cuhe: A homomorphic encryption accelerator library,'' in
  \emph{International Conference on Cryptography and Information Security in
  the Balkans}.\hskip 1em plus 0.5em minus 0.4em\relax Springer, 2015, pp.
  169--186.

\bibitem{nufhe}
\BIBentryALTinterwordspacing
nucypher. (2020) Nufhe, a gpu-powered torus fhe implementation. [Online].
  Available: \url{https://github.com/nucypher/nufhe}
\BIBentrySTDinterwordspacing

\bibitem{riazi2020heax}
M.~S. Riazi, K.~Laine, B.~Pelton, and W.~Dai, ``Heax: An architecture for
  computing on encrypted data,'' in \emph{Proceedings of the Twenty-Fifth
  International Conference on Architectural Support for Programming Languages
  and Operating Systems}, 2020, pp. 1295--1309.

\bibitem{agrawal2022fab}
R.~Agrawal, L.~de~Castro, G.~Yang, C.~Juvekar, R.~Yazicigil, A.~Chandrakasan,
  V.~Vaikuntanathan, and A.~Joshi, ``Fab: An fpga-based accelerator for
  bootstrappable fully homomorphic encryption,'' \emph{arXiv preprint
  arXiv:2207.11872}, 2022.

\bibitem{yang2023poseidon}
Y.~Yang, H.~Zhang, S.~Fan, H.~Lu, M.~Zhang, and X.~Li, ``Poseidon: Practical
  homomorphic encryption accelerator,'' in \emph{2023 IEEE International
  Symposium on High-Performance Computer Architecture (HPCA)}.\hskip 1em plus
  0.5em minus 0.4em\relax IEEE, 2023, pp. 870--881.

\bibitem{gener2021fpga}
S.~Gener, P.~Newton, D.~Tan, S.~Richelson, G.~Lemieux, and P.~Brisk, ``An
  fpga-based programmable vector engine for fast fully homomorphic encryption
  over the torus,'' in \emph{SPSL: Secure and Private Systems for Machine
  Learning (ISCA Workshop)}, 2021.

\bibitem{samardzic2021f1}
N.~Samardzic, A.~Feldmann, A.~Krastev, S.~Devadas, R.~Dreslinski, C.~Peikert,
  and D.~Sanchez, ``F1: A fast and programmable accelerator for fully
  homomorphic encryption,'' in \emph{MICRO-54: 54th Annual IEEE/ACM
  International Symposium on Microarchitecture}, 2021, pp. 238--252.

\bibitem{samardzic2022craterlake}
N.~Samardzic, A.~Feldmann, A.~Krastev, N.~Manohar, N.~Genise, S.~Devadas,
  K.~Eldefrawy, C.~Peikert, and D.~Sanchez, ``Craterlake: a hardware
  accelerator for efficient unbounded computation on encrypted data.'' in
  \emph{ISCA}, 2022, pp. 173--187.

\bibitem{kim2022bts}
S.~Kim, J.~Kim, M.~J. Kim, W.~Jung, J.~Kim, M.~Rhu, and J.~H. Ahn, ``Bts: An
  accelerator for bootstrappable fully homomorphic encryption,'' in
  \emph{Proceedings of the 49th Annual International Symposium on Computer
  Architecture}, 2022, pp. 711--725.

\bibitem{kim2022ark}
J.~Kim, G.~Lee, S.~Kim, G.~Sohn, J.~Kim, M.~Rhu, and J.~H. Ahn, ``Ark: Fully
  homomorphic encryption accelerator with runtime data generation and
  inter-operation key reuse,'' \emph{arXiv preprint arXiv:2205.00922}, 2022.

\bibitem{cheon2017homomorphic}
J.~H. Cheon, A.~Kim, M.~Kim, and Y.~Song, ``Homomorphic encryption for
  arithmetic of approximate numbers,'' in \emph{International conference on the
  theory and application of cryptology and information security}.\hskip 1em
  plus 0.5em minus 0.4em\relax Springer, 2017, pp. 409--437.

\bibitem{chillotti2020tfhe}
I.~Chillotti, N.~Gama, M.~Georgieva, and M.~Izabach{\`e}ne, ``Tfhe: fast fully
  homomorphic encryption over the torus,'' \emph{Journal of Cryptology},
  vol.~33, no.~1, pp. 34--91, 2020.

\bibitem{jiang2022matcha}
L.~Jiang, Q.~Lou, and N.~Joshi, ``Matcha: A fast and energy-efficient
  accelerator for fully homomorphic encryption over the torus,'' \emph{arXiv
  preprint arXiv:2202.08814}, 2022.

\bibitem{ye2022HPEC}
T.~Ye, R.~Kannan, and V.~K. Prasanna, ``Fpga acceleration of fully homomorphic
  encryption over the torus,'' in \emph{2022 IEEE High Performance Extreme
  Computing Conference (HPEC)}, 2022, pp. 1--7.

\bibitem{nam22nbit}
\BIBentryALTinterwordspacing
K.~Nam, H.~Oh, H.~Moon, and Y.~Paek, ``Accelerating n-bit operations over tfhe
  on commodity cpu-fpga,'' in \emph{Proceedings of the 41st IEEE/ACM
  International Conference on Computer-Aided Design}, ser. ICCAD '22.\hskip 1em
  plus 0.5em minus 0.4em\relax New York, NY, USA: Association for Computing
  Machinery, 2022. [Online]. Available:
  \url{https://doi.org/10.1145/3508352.3549413}
\BIBentrySTDinterwordspacing

\bibitem{regev2009lattices}
O.~Regev, ``On lattices, learning with errors, random linear codes, and
  cryptography,'' \emph{Journal of the ACM (JACM)}, vol.~56, no.~6, pp. 1--40,
  2009.

\bibitem{bernstein2009introduction}
D.~J. Bernstein, ``Introduction to post-quantum cryptography,'' in
  \emph{Post-quantum cryptography}.\hskip 1em plus 0.5em minus 0.4em\relax
  Springer, 2009, pp. 1--14.

\bibitem{khot2005hardness}
S.~Khot, ``Hardness of approximating the shortest vector problem in lattices,''
  \emph{Journal of the ACM (JACM)}, vol.~52, no.~5, pp. 789--808, 2005.

\bibitem{gentry2009fullybook}
C.~Gentry, \emph{A fully homomorphic encryption scheme}.\hskip 1em plus 0.5em
  minus 0.4em\relax Stanford university, 2009.

\bibitem{gentry2011implementing}
C.~Gentry and S.~Halevi, ``Implementing gentry’s fully-homomorphic encryption
  scheme,'' in \emph{Annual international conference on the theory and
  applications of cryptographic techniques}.\hskip 1em plus 0.5em minus
  0.4em\relax Springer, 2011, pp. 129--148.

\bibitem{brakerski2014leveled}
Z.~Brakerski, C.~Gentry, and V.~Vaikuntanathan, ``(leveled) fully homomorphic
  encryption without bootstrapping,'' \emph{ACM Transactions on Computation
  Theory (TOCT)}, vol.~6, no.~3, pp. 1--36, 2014.

\bibitem{fan2012somewhat}
J.~Fan and F.~Vercauteren, ``Somewhat practical fully homomorphic encryption,''
  \emph{Cryptology ePrint Archive}, 2012.

\bibitem{gentry2013homomorphic}
C.~Gentry, A.~Sahai, and B.~Waters, ``Homomorphic encryption from learning with
  errors: Conceptually-simpler, asymptotically-faster, attribute-based,'' in
  \emph{Annual Cryptology Conference}.\hskip 1em plus 0.5em minus 0.4em\relax
  Springer, 2013, pp. 75--92.

\bibitem{ducas2015fhew}
L.~Ducas and D.~Micciancio, ``Fhew: bootstrapping homomorphic encryption in
  less than a second,'' in \emph{Annual international conference on the theory
  and applications of cryptographic techniques}.\hskip 1em plus 0.5em minus
  0.4em\relax Springer, 2015, pp. 617--640.

\bibitem{chillotti2016faster}
I.~Chillotti, N.~Gama, M.~Georgieva, and M.~Izabachene, ``Faster fully
  homomorphic encryption: Bootstrapping in less than 0.1 seconds,'' in
  \emph{international conference on the theory and application of cryptology
  and information security}.\hskip 1em plus 0.5em minus 0.4em\relax Springer,
  2016, pp. 3--33.

\bibitem{falcetta2022privacy}
A.~Falcetta and M.~Roveri, ``Privacy-preserving deep learning with homomorphic
  encryption: An introduction,'' \emph{IEEE Computational Intelligence
  Magazine}, vol.~17, no.~3, pp. 14--25, 2022.

\bibitem{podschwadt2022survey}
R.~Podschwadt, D.~Takabi, P.~Hu, M.~H. Rafiei, and Z.~Cai, ``A survey of deep
  learning architectures for privacy-preserving machine learning with fully
  homomorphic encryption,'' \emph{IEEE Access}, 2022.

\bibitem{marcolla2022survey}
C.~Marcolla, V.~Sucasas, M.~Manzano, R.~Bassoli, F.~H. Fitzek, and N.~Aaraj,
  ``Survey on fully homomorphic encryption, theory, and applications,''
  \emph{Proceedings of the IEEE}, 2022.

\bibitem{chillotti2021programmable}
I.~Chillotti, M.~Joye, and P.~Paillier, ``Programmable bootstrapping enables
  efficient homomorphic inference of deep neural networks,'' in
  \emph{International Symposium on Cyber Security Cryptography and Machine
  Learning}.\hskip 1em plus 0.5em minus 0.4em\relax Springer, 2021, pp. 1--19.

\bibitem{chillotti2016homomorphic}
I.~Chillotti, N.~Gama, M.~Georgieva, and M.~Izabach{\`e}ne, ``A homomorphic lwe
  based e-voting scheme,'' in \emph{Post-Quantum Cryptography}.\hskip 1em plus
  0.5em minus 0.4em\relax Springer, 2016, pp. 245--265.

\bibitem{chillotti2017faster}
------, ``Faster packed homomorphic operations and efficient circuit
  bootstrapping for tfhe,'' in \emph{International Conference on the Theory and
  Application of Cryptology and Information Security}.\hskip 1em plus 0.5em
  minus 0.4em\relax Springer, 2017, pp. 377--408.

\bibitem{chillotti2021improved}
I.~Chillotti, D.~Ligier, J.-B. Orfila, and S.~Tap, ``Improved programmable
  bootstrapping with larger precision and efficient arithmetic circuits for
  tfhe,'' in \emph{International Conference on the Theory and Application of
  Cryptology and Information Security}.\hskip 1em plus 0.5em minus 0.4em\relax
  Springer, 2021, pp. 670--699.

\bibitem{chillotti2022scooby}
I.~Chillotti, E.~Orsini, P.~Scholl, N.~P. Smart, and B.~Van~Leeuwen, ``Scooby:
  Improved multi-party homomorphic secret sharing based on fhe,''
  \emph{Cryptology ePrint Archive}, 2022.

\bibitem{lou2019she}
Q.~Lou and L.~Jiang, ``She: A fast and accurate privacy-preserving deep neural
  network via leveled tfhe and logarithmic data representation,'' \emph{arXiv:
  1906.00148}, 2019.

\bibitem{stoian23dnntfhe}
\BIBentryALTinterwordspacing
A.~Stoian, J.~Frery, R.~Bredehoft, L.~Montero, C.~Kherfallah, and
  B.~Chevallier-Mames, ``Deep neural networks for encrypted inference with
  tfhe,'' Cryptology ePrint Archive, Paper 2023/257, 2023,
  \url{https://eprint.iacr.org/2023/257}. [Online]. Available:
  \url{https://eprint.iacr.org/2023/257}
\BIBentrySTDinterwordspacing

\bibitem{frery23pptree}
\BIBentryALTinterwordspacing
J.~Frery, A.~Stoian, R.~Bredehoft, L.~Montero, C.~Kherfallah,
  B.~Chevallier-Mames, and A.~Meyre, ``Privacy-preserving tree-based inference
  with fully homomorphic encryption,'' Cryptology ePrint Archive, Paper
  2023/258, 2023, \url{https://eprint.iacr.org/2023/258}. [Online]. Available:
  \url{https://eprint.iacr.org/2023/258}
\BIBentrySTDinterwordspacing

\bibitem{matsuoka2021virtual}
K.~Matsuoka, R.~Banno, N.~Matsumoto, T.~Sato, and S.~Bian, ``Virtual secure
  platform: A $\{$Five-Stage$\}$ pipeline processor over $\{$TFHE$\}$,'' in
  \emph{30th USENIX Security Symposium (USENIX Security 21)}, 2021, pp.
  4007--4024.

\bibitem{joye2022sok}
M.~Joye, ``Sok: Fully homomorphic encryption over the [discretized] torus,''
  \emph{IACR Transactions on Cryptographic Hardware and Embedded Systems}, pp.
  661--692, 2022.

\bibitem{chillotti2020concrete}
I.~Chillotti, M.~Joye, D.~Ligier, J.-B. Orfila, and S.~Tap, ``Concrete:
  Concrete operates on ciphertexts rapidly by extending tfhe,'' in \emph{WAHC
  2020--8th Workshop on Encrypted Computing \& Applied Homomorphic
  Cryptography}, vol.~15, 2020.

\bibitem{garrido2021survey}
M.~Garrido, ``A survey on pipelined fft hardware architectures,'' \emph{Journal
  of Signal Processing Systems}, pp. 1--20, 2021.

\bibitem{rabiner1979use}
L.~Rabiner, ``On the use of symmetry in fft computation,'' \emph{IEEE
  Transactions on Acoustics, Speech, and Signal Processing}, vol.~27, no.~3,
  pp. 233--239, 1979.

\bibitem{guo1998quick}
H.~Guo, G.~A. Sitton, and C.~S. Burrus, ``The quick fourier transform: an fft
  based on symmetries,'' \emph{IEEE transactions on signal processing},
  vol.~46, no.~2, pp. 335--341, 1998.

\bibitem{klemsa2021fast}
J.~Klemsa, ``Fast and error-free negacyclic integer convolution using extended
  fourier transform,'' in \emph{International Symposium on Cyber Security
  Cryptography and Machine Learning}.\hskip 1em plus 0.5em minus 0.4em\relax
  Springer, 2021, pp. 282--300.

\bibitem{jedecHBM}
J.~S. S.~T. ASSOCIATION, ``{JEDEC Standard: HBM DRAM},'' JECDC, Tech. Rep., 10
  2013.

\bibitem{de2021does}
L.~de~Castro, R.~Agrawal, R.~Yazicigil, A.~Chandrakasan, V.~Vaikuntanathan,
  C.~Juvekar, and A.~Joshi, ``Does fully homomorphic encryption need compute
  acceleration?'' \emph{arXiv preprint arXiv:2112.06396}, 2021.

\bibitem{bourse2018fast}
F.~Bourse, M.~Minelli, M.~Minihold, and P.~Paillier, ``Fast homomorphic
  evaluation of deep discretized neural networks,'' in \emph{Annual
  International Cryptology Conference}.\hskip 1em plus 0.5em minus 0.4em\relax
  Springer, 2018, pp. 483--512.

\end{thebibliography}
%%%%%%%%%%%%%%%%%%%%%%%%%%%%%%%%%%%%

\end{document}